\documentclass[paper]{JHEP3}
\pdfoutput=1
\usepackage{amsmath,amssymb,amsthm,amscd,graphicx}
\usepackage{psfrag}
\input epsf.sty
\theoremstyle{definition}

\newcommand{\CC}{{\cal C}}

\newcommand{\CF}{{\cal F}}

\newcommand{\CH}{{\cal H}}

\newcommand{\CO}{{\cal O}}

\def\IZ{{\mathbb Z}}
\def\IF{{\mathbb F}}
\def\IR{{\mathbb R}}
\def\IC{{\mathbb C}}
\def\IP{{\mathbb P}}

\newcommand{\tr}{{\rm Tr}}
\newcommand{\re}{{\rm e}}
\newcommand{\ri}{{\rm i}}
\newcommand{\rd}{{\rm d}}
\newcommand{\mx}{{\mathsf{x}}}
\newcommand{\my}{{\mathsf{y}}}
\newcommand{\mb}{{\mathsf{b}}}

\newcommand{\im}{{\mathsf{i}}}
\newcommand{\mg}{{\mathsf{g}}}

\newcommand{\fad}{\operatorname{\Phi}_{\mathsf{b}}}

\newcommand{\fadm}{\operatorname{\Phi}_{1/\mathsf{b}}}

\newcommand{\mypsi}[2]{\operatorname{\Psi}_{#1,#2}}

\newcommand{\be}{\begin{equation}}
\newcommand{\ee}{\end{equation}}
\newcommand{\ba}{\begin{aligned}}
\newcommand{\ea}{\end{aligned}}
\newcommand{\ben}{\begin{eqnarray}\displaystyle}
\newcommand{\een}{\end{eqnarray}}

\newcommand{\sectiono}[1]{\section{#1}\setcounter{equation}{0}}

\newdimen\tableauside\tableauside=1.0ex
\newdimen\tableaurule\tableaurule=0.4pt
\newdimen\tableaustep
\def\phantomhrule#1{\hbox{\vbox to0pt{\hrule height\tableaurule width#1\vss}}}
\def\phantomvrule#1{\vbox{\hbox to0pt{\vrule width\tableaurule height#1\hss}}}
\def\sqr{\vbox{%
  \phantomhrule\tableaustep
  \hbox{\phantomvrule\tableaustep\kern\tableaustep\phantomvrule\tableaustep}%
  \hbox{\vbox{\phantomhrule\tableauside}\kern-\tableaurule}}}
\def\squares#1{\hbox{\count0=#1\noindent\loop\sqr
  \advance\count0 by-1 \ifnum\count0>0\repeat}}
\def\tableau#1{\vcenter{\offinterlineskip
  \tableaustep=\tableauside\advance\tableaustep by-\tableaurule
  \kern\normallineskip\hbox
    {\kern\normallineskip\vbox
      {\gettableau#1 0 }%
     \kern\normallineskip\kern\tableaurule}%
  \kern\normallineskip\kern\tableaurule}}
\def\gettableau#1{\ifnum#1=0\let\next=\null\else
\squares{#1}\let\next=\gettableau\fi\next}

\newcommand{\figref}[1]{Fig.~\protect\ref{#1}}
\title{\huge{Matrix models from operators and topological strings}}

\author{
Marcos Mari\~no and Szabolcs Zakany\\
D\'epartement de Physique Th\'eorique et Section de Math\'ematiques,\\
Universit\'e de Gen\`eve, Gen\`eve, CH-1211 Switzerland
\\
\\
\email{marcos.marino@unige.ch, szabolcs.zakany@unige.ch}
}

\abstract{We propose a new family of matrix models whose $1/N$ expansion captures the all-genus topological string on toric Calabi--Yau threefolds. 
These matrix models are constructed from the trace class operators appearing in the quantization of the corresponding mirror curves. The fact that they provide a 
non-perturbative realization of the (standard) topological string follows from a recent conjecture connecting the spectral 
properties of these operators, to the enumerative invariants of the underlying Calabi--Yau threefolds. We study in detail the resulting matrix models for some simple 
geometries, like local $\IP^2$ and local $\IF_2$, and 
we verify that their weak 't Hooft coupling expansion reproduces the topological string free energies near the conifold singularity. 
These matrix models are formally similar to those appearing in the Fermi-gas formulation of Chern--Simons--matter theories, and their $1/N$ expansion receives non-perturbative 
corrections determined by the Nekrasov--Shatashvili limit of the refined topological string.}

\begin{document}
 
 \sectiono{Introduction}
 
 One of the most surprising aspects of string theory is that, in some circumstances, it can be described by very simple quantum systems. For example, 
 non-critical (super)strings can be formulated in terms of double-scaled matrix models or matrix quantum mechanics.  
 In some cases, these equivalent descriptions provide as well a non-perturbative definition of the corresponding 
 string theory models. More recently, some simple quantities in fully-fledged superstring theory, like partition functions, have been also 
 expressed in terms of matrix integrals, by combining the AdS/CFT correspondence with supersymmetric localization. 
 
 It has been suspected for a long time that 
 topological strings on Calabi--Yau (CY) manifolds should be also described by simple quantum models. It was found in \cite{dv} that 
 the type B topological string on a special class of non-compact B-model geometries has such a description, in terms of conventional matrix models. 
The CY backgrounds considered in \cite{dv} are useful from the point of view of engineering supersymmetric gauge theories, 
but they have no mirror geometries and no enumerative content. 
It was later found that {\it bona fide} topological strings on $A_n$ fibrations over $\IP^1$ can be described by Chern--Simons matrix models \cite{mmcs}, as a 
consequence of the Gopakumar--Vafa large $N$ duality \cite{gv} and its generalizations \cite{akmv}. 

Recently, a correspondence has been proposed between topological strings on 
toric CY threefolds, and the spectral theory of operators arising in the quantization 
of their mirror curves \cite{ghm}. The idea that the topological string free energies could 
emerge from the quantization of mirror curves was first proposed in \cite{adkmv}. In \cite{acdkv,mirmor},
 building upon the work of \cite{ns}, it was shown that a perturbative 
treatment of the quantum mirror curve leads to the Nekrasov--Shatashvili (NS) limit of the refined topological string. However, 
building on the study of matrix models for Chern--Simons--matter theories \cite{kwy} and their 
AdS/CFT duals \cite{abjm} (see \cite{mm-lectures} for a review), it was pointed out in \cite{km} that
the {\it standard} topological string {\it also} emerges from the quantum curve, once non-perturbative corrections are taken into account. 

The proposal of \cite{ghm} incorporates all these ingredients in an exact treatment of the quantum curve. According 
to \cite{ghm}, to each mirror curve of a toric Calabi--Yau threefold, one can associate a positive, 
trace class operator on $L^2(\IR)$. This was rigorously proved 
for a large number of geometries in \cite{kas-mar}. Therefore, these operators have a positive, 
discrete spectrum, and their Fredholm or spectral determinants are well defined. In \cite{ghm}, an explicit 
formula for these spectral determinants was conjectured, involving both the 
NS limit of the refined topological string and the conventional topological string. 
The conjectural, exact formula of \cite{ghm} leads in addition to an exact 
quantization condition determining the spectrum, which generalizes previous studies of the spectral problem 
\cite{km,hw, kallen, fhw} and is conceptually similar to other exact quantization conditions appearing in Quantum Mechanics 
(see for example \cite{zjj}). This establishes a novel and precise connection 
between spectral theory and mirror symmetry. 

\begin{figure}
\center
\includegraphics[height=6cm]{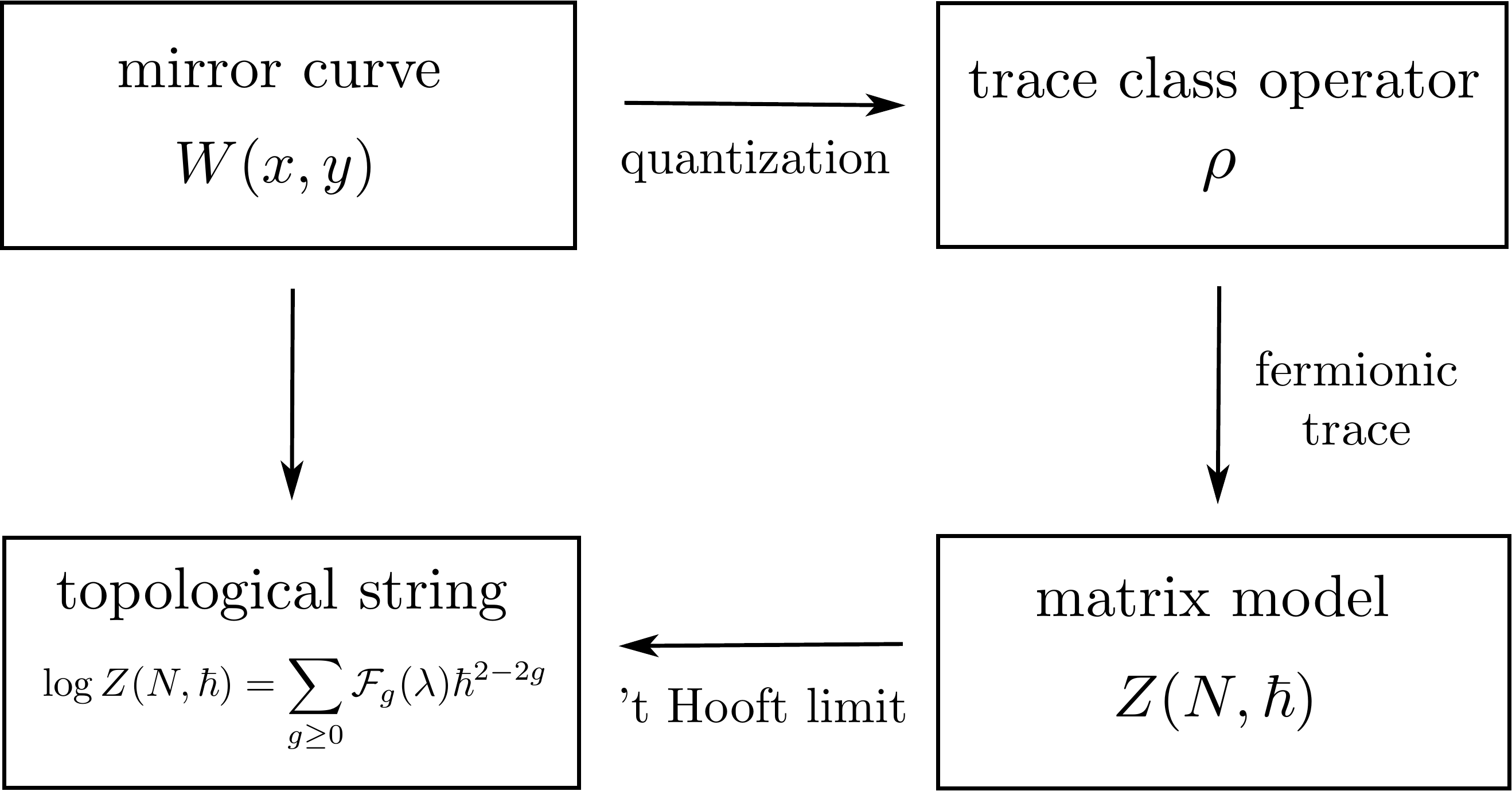} 
\caption{Given a toric Calabi--Yau threefold, the quantization of its mirror curve leads to a trace class operator $\rho$. 
The standard topological string free energy is obtained as the 't Hooft limit of its fermionic traces $Z(N, \hbar)$.}
\label{rels}
\end{figure}
One of the consequences of the correspondence of \cite{ghm} is that the conventional topological string free 
energy (at all genera) appears as a 't Hooft limit 
of the spectral determinant. A very useful way to encode the information in the spectral determinant is in terms of the 
so-called fermionic spectral traces $Z(N, \hbar)$ (see section \ref{gen-as} for precise definitions). As we will show in detail, 
these traces have a natural matrix model representation. It then follows from the conjecture of \cite{ghm} that the 't Hooft limit of this matrix model, 
\be
\label{thooftlimit}
N \rightarrow \infty, \qquad \hbar \rightarrow \infty, \qquad {N \over \hbar}=\lambda \, \, \, {\text{fixed}}, 
\ee
is given by the asymptotic expansion
\be
\label{free-expansion}
\log Z(N, \hbar)= \sum_{g\ge 0} \CF_g(\lambda) \hbar^{2-2g}, 
\ee
where $\CF_g(\lambda)$ are the standard topological string free energies, and the 't Hooft parameter 
$\lambda$ is a flat coordinate for the CY moduli space\footnote{In this paper, as in \cite{ghm}, we will focus on local del Pezzo Calabi--Yau's, 
where there is a single ``true" modulus, and correspondingly a single 't Hooft parameter.}. 
Therefore, the conjecture of \cite{ghm}, together with the representation of $Z(N, \hbar)$ 
in terms of a matrix integral, gives a matrix model for topological strings on toric Calabi--Yau threefolds. 
The construction is summarized in \figref{rels}. We should add that the resulting matrix model 
is a convergent one, i.e. the matrix integral is well-defined, as a consequence of the operator being of trace class. 

As explained in \cite{ghm}, the correspondence between spectral theory and mirror symmetry provides a 
rigorous, non-perturbative completion of the topological string, in the sense that the genus expansion of 
its free energy, which is known to be a divergent series, is realized as the asymptotic expansion of a well-defined function. 
The implementation of the correspondence of \cite{ghm} that we are presenting here, in terms of matrix models, 
makes this point particularly clear: the quantity 
$Z(N, \hbar)$ is manifestly well-defined for any positive integer $N$ and any real $\hbar$, since it is defined by the spectral theory of a trace class operator (it can be 
also analytically continued to complex $N$, as first explained in \cite{cgm}, and it is likely that it has an analytic continuation to complex values of $\hbar$). 
The 't Hooft expansion of $Z(N, \hbar)$ is given exactly by the genus expansion of the topological string free energy. 
However, there are non-perturbative corrections to the 't Hooft expansion, due to large $N$ instantons, 
which are also predicted by the conjecture in \cite{ghm}, and they are encoded in the NS limit of the refined topological string.    

The fact that the matrix model representing $Z(N, \hbar)$ leads 
to the topological string free energies is {\it not} at all obvious. On the contrary, 
it is a highly non-trivial prediction of the conjecture of \cite{ghm}. The 't Hooft limit of $Z(N, \hbar)$ probes
 the {\it strong coupling limit} of the spectral problem (since $\hbar$ is large), and in particular 
 the non-perturbative instanton corrections to the perturbative WKB expansion. 
 For this reason, in this paper we will perform detailed calculations in some examples to verify that, indeed, the't Hooft expansion of the 
 matrix model defined by the trace class operator gives 
the topological string free energies. This constitutes an {\it analytic} test of the instanton corrections postulated in \cite{ghm}. 

As we mentioned above, the large $N$ duality of Gopakumar--Vafa \cite{gv} and its generalizations \cite{akmv} 
provide a matrix model representation of the free energies 
of topological strings in certain geometries, as it was tested in \cite{akmv,hy,hoy}. 
Although in this paper we focus on local del Pezzo geometries, our matrix models are potentially 
valid for any toric geometry, in contrast to the duality of \cite{gv,akmv}. For example, 
we will study in detail a matrix model for local $\IP^2$, which has no counterpart in the framework of 
\cite{gv,akmv}. It would be interesting to understand the relationship between the 
matrix models for topological strings obtained in \cite{akmv} and the ones described here. 

Matrix models describing topological strings on more general backgrounds were also proposed in for example \cite{eynard,ks,sul,ekm1,ekm2}. We 
should note that these models are 
very different from the ones we construct here: first of all, they are engineered {\it ab initio} to reproduce formally the topological string free energies; 
in contrast, our matrix models are defined by the trace class operators obtained in the quantization of the mirror curve, and the 
fact that they lead to the correct topological string free energies is a consequence of the non-trivial conjecture of \cite{ghm}. Second, in 
the models of \cite{eynard,ks,sul,ekm1,ekm2}, the rank of the matrix $N$ 
plays an auxiliary role, while in our case it is a flat coordinate for the Calabi--Yau, as in other large $N$ dualities. 
Third, the matrix models in \cite{eynard,ks,sul,ekm1,ekm2} are often formal (i.e. not convergent) 
and therefore can not define a non-perturbative completion of the theory; our matrix models are convergent and lead to a non-perturbative completion. 

This paper is organized as follows. In section 2 we review elementary aspects of trace class operators on $L^2(\IR)$ and we note that their fermionic traces have matrix model-like 
representations. We then focus on the operators coming from quantized mirror curves, and write down explicit matrix models 
for some of them, including the ones relevant for local $\IP^2$ and local $\IF_2$. These models can be studied in the 't Hooft expansion, and we compute their 
weakly coupled expansion to the very first orders. In section 3 we review the conjecture of \cite{ghm} and we spell out in detail its prediction for the 't Hooft expansion of the 
fermionic traces. Then, we test the conjecture in detail for the matrix models describing local $\IP^2$ and local $\IF_2$. 
We conclude in section 4 and we list some open problems for the future. In the Appendix, we list some results for 
the weakly coupled expansion of the matrix model free energies.

\sectiono{From operators to matrix models}

\subsection{General aspects}
\label{gen-as}

Let us then begin with some general aspects of operator theory. Let $\rho$ be a positive-definite, trace class operator on the Hilbert space $\CH=L^2(\IR)$, 
depending on a real parameter $\hbar$. Since its eigenvalues are discrete and positive, we will 
denote them by $\re^{-E_n}$, where $n=0, 1,\cdots$. Due to the trace class property, all the {\it spectral traces} of $\rho$
\be
\label{s-traces}
Z_\ell = \tr_\CH \rho^\ell= \sum_{n \ge 0} \re^{-\ell E_n}, \qquad \ell=1, 2, \cdots, 
\ee
exist. One can also define the {\it fermionic spectral traces} as 
\be
Z(N, \hbar) = \tr_{\Lambda^N\left(\CH \right)} \left(\Lambda^N(\rho)\right), \qquad N=1, 2, \cdots, 
\ee
where the operator $\Lambda^N(\rho)$ is defined by $\rho^{\otimes N}$ acting on $\Lambda^N\left(\CH \right)$ (see for example \cite{simon} for a 
review of these constructions)\footnote{The terms ``spectral trace" and ``fermionic spectral trace" do not seem to be standard, 
but since these objects are of paramount importance in our 
construction, we had to find a name for them.}. The fermionic spectral traces are related to the standard spectral traces (\ref{s-traces}) by the equation 
\be
\label{conjclasses}
Z(N, \hbar) =\sum_{\{ m_\ell \}} {}^{'}\prod_\ell   {(-1)^{(\ell-1)m_\ell } Z_\ell^{m_\ell} \over m_\ell! \ell^{m_\ell}},
\ee
where the $ {}^{'}$ means that the sum is over the integers $m_\ell$ satisfying the constraint
\be
\label{Ncons}
\sum_\ell \ell m_\ell=N.
\ee
The {\it Fredholm} or {\it spectral determinant} of $\rho$ can be defined as 
\be
\label{f-det}
\Xi(\kappa, \hbar)= {\rm det}(1+ \kappa  \rho)= \prod_{n=0}^\infty \left(1+ \kappa\re^{-E_n} \right),
\ee
or, equivalently, by the expansion around $\kappa=0$, 
\be
\Xi(\kappa, \hbar)=1+\sum_{N=1}^\infty Z(N, \hbar) \kappa^N, 
\ee
and is an entire function of $\kappa$ \cite{simon}. Note that, if $\rho$ is interpreted as a one-particle thermal density operator, the fermionic trace 
$Z(N, \hbar)$ is the canonical partition function for an ideal Fermi gas 
of $N$ particles. It then follows that $Z(N, \hbar)$ has the matrix-model-like representation 
\be
\label{znmm}
Z(N, \hbar)= {1 \over N!} \sum_{\sigma  \in S_N} (-1)^{\epsilon(\sigma)}  \int  \rd ^N x \prod_i \rho(x_i, x_{\sigma(i)})
= {1 \over N!}  \int  \rd ^N x \, {\rm det}\left( \rho(x_i, x_j) \right). 
\ee
In this equation, $\rho(x_1, x_2)$ is the kernel of the operator $\rho$, 
\be
\rho(x_1, x_2) = \langle x_1 |\rho |x_2 \rangle, 
\ee
$S_N$ is the permutation group of $N$ elements, and $\epsilon (\sigma)$ is the signature of a permutation $\sigma \in S_N$. The above formula 
encapsulates the fermionic nature of $Z(N, \hbar)$. 

At this stage, calling (\ref{znmm}) a matrix model might seem excessive. However, 
the integrand of (\ref{znmm}) has an essential property, typical of the integrand of a matrix model in the eigenvalue representation: it vanishes whenever 
$x_i=x_j$. In standard matrix models, this is an indication of eigenvalue repulsion. To further understand 
this parallelism, note that, when solved in terms of orthogonal polynomials, 
the partition function of a Hermitian matrix model is given by the expression
\be
\label{znop}
Z_N= {1 \over N!}  \int  \rd ^N x \, {\rm det}\left(K_N(x_i, x_j) \right), 
\ee
where the kernel $K_N(x, y)$ can be written in terms of the first $N$ orthogonal polynomials. This is very similar to (\ref{znmm}). A crucial difference though 
between (\ref{znmm}) and the 
more familiar expression (\ref{znop}) is that, in (\ref{znmm}), the kernel does not depend on $N$, and the resulting matrix models 
are quite special. Matrix models of the form (\ref{znmm}), 
coming from integral kernels of operators, have been considered before in for example \cite{zamo,kostov}, and they have been recently studied from a 
more general point of view in \cite{grassi-marino}, where the notion of ``M-theoretic matrix model" was introduced.

As we will see in the next sections, in the case of trace class operators 
obtained by quantization of mirror curves, the matrix model (\ref{znmm}) is closely related to Chern--Simons matrix models \cite{mmcs} and to the 
matrix models appearing in the localization of Chern--Simons--matter theories \cite{kwy,hhl,jafferis} (see \cite{mm-lectures} for a review). 
In particular, the resulting matrix models are M-theoretic, in the sense of \cite{grassi-marino}: they 
have a well-defined 't Hooft expansion in the limit (\ref{thooftlimit}), but they also have a well-defined M-theory limit, in which $N$ is large but $\hbar$ is fixed. 

\subsection{Operators from mirror curves and matrix models}

In \cite{ghm}, building on previous work on the quantization of mirror curves, it was postulated that, given a mirror curve to a toric CY threefold, one can 
quantize it to obtain a trace class operator $\rho$. As in \cite{ghm,kas-mar}, we will focus on toric (almost) del Pezzo CY threefolds, which are 
defined as the total space of the anti-canonical bundle on a toric (almost) del Pezzo surface $S$,
\be
\label{dP}
X=\CO(-K_S) \rightarrow S. 
\ee
The mirror curve depends on $k$ complex moduli $z_\alpha$, $\alpha=1, \cdots, k$ and is of genus one. This means that the complex moduli 
involve a ``true" geometric modulus $\tilde u$ and a set of ``mass" parameters $\xi_i$, $i=1, \cdots, r$, where $r$ depends on the geometry 
under consideration \cite{hkp,hkrs}. The mirror curves can be put in the form
\be
\label{ex-W}
W(\re^x, \re^y)= \CO_S(x,y)+ \tilde u=0,  
\ee
where $\CO_S(x,y)$ has the form
\be
\label{coxp}
 \CO_S (x,y)=\sum_{i=1}^{k+2} \exp\left( \nu^{(i)}_1 x+  \nu^{(i)}_2 y + f_i(\xi_j) \right), 
 \ee
and $f_i(\xi_j)$ are suitable functions of the parameters $\xi_j$. The vectors $\nu^{(i)}_{1,2}$ can be obtained from the fan defining the toric 
CY threefold. 

In this paper we will be particularly interested in local $\IP^2$, where the function $\CO_S(x,y)$ is given by, 
\be
\label{lp2}
\CO_{\IP^2} \left(x, y \right)= \re^{ x} + \re^{y} + \re^{- x- y}, 
\ee
and local $\IF_2$, where it is given by, 
\be
\label{f2}
\CO_{\IF_2} \left(x, y \right)= \re^{ x} + \re^{y} + \re^{- 2 x- y} + \xi \re^{-x}. 
\ee
To quantize the mirror curve (\ref{ex-W}), we promote $x$, $y$ to self-adjoint Heisenberg 
operators $\mathsf{x}$, $\mathsf{y}$ satisfying the commutation relation 
\be
[\mathsf{x}, \mathsf{y}]=\im\hbar. 
\ee
This promotes $\CO_S(x,y)$ to an operator, which 
will be denoted by $\mathsf{O}_S$ (possible ordering ambiguities are resolved by requiring the resulting operator to be self-adjoint). 
As conjectured in \cite{ghm} and proved in \cite{kas-mar} for 
many geometries, the inverse operator
\be
\rho_S=\mathsf{O}^{-1}_S
\ee
is positive-definite and of trace class. By the construction explained in the previous section, the fermionic traces 
of this operator, which we will denote by $Z_S(N, \hbar)$, have an integral 
representation in terms of the kernel of the operator $\rho_S$. Therefore, in order to write down the 
matrix model (\ref{znmm}), we need an explicit expression for this kernel. In general, 
obtaining such an expression is not easy. However, in \cite{kas-mar}, this problem was solved for three-term operators of the form 
\be
\mathsf{O}_{m,n}=\re^{\mx}+ \re^{\my}+ \re^{-m \mx - n \my}. 
\ee
Geometrically, this operator corresponds to the anti-canonical bundle of the weigthed projective space $\IP(1,n,m)$. 
When $n=1$, these geometries can be constructed as partial 
blow-ups of the orbifold $\IC^3/\IZ_{m+2}$, where the orbifold action has the weights $(1,1, m)$ \cite{bc}. 
In principle, more general geometries can be obtained by perturbing the operator $\mathsf{O}_{m,n}$ 
appropriately. Note that the case $m=n=1$ gives local $\IP^2$, while $m=2$, $n=1$ gives the $\xi=0$ limit of local $\IF_2$, i.e. a partial blowdown of the local $\IF_2$ geometry.
Let us define
\be
\rho_{m,n}= \mathsf{O}_{m,n}^{-1}. 
\ee
As shown in \cite{kas-mar}, the kernel of $\rho_{m,n}$ involves in a crucial way Faddeev's quantum dilogarithm $ \fad(x)$ \cite{faddeev,fk}
(in this paper, we use the notations of \cite{kas-mar} for this function). We will also need 
the function 
\be
\label{mypsi-def}
\mypsi{a}{c}(x)= \frac{\re^{2\pi ax}}{\fad(x-\im(a+c))}, 
\ee
which behaves at large $|x|$ as
\be
\mypsi{a}{c}(x) \approx \begin{cases} \re^{-2 \pi c x}, & x \rightarrow \infty,\\
\re^{2 \pi a x}, & x \rightarrow -\infty. \end{cases}
\ee
In \cite{kas-mar} it was shown that, in terms of an appropriate variable $p$, related to $x,y$ by a linear canonical transformation, one has 
\be
\label{rmn-ker}
\rho_{m,n}(p_1,p_2) = \langle p_1| \rho_{m,n} |p_2 \rangle = \frac{\overline{\mypsi{a}{c}(p_1)} \mypsi{a}{c}(p_2)}{2 \mb \cosh \left (\frac{\pi}{\mb}(p_1-p_2)+{\im \pi \over \mb} \left(a+c-nc\right) \right)}. 
\ee
In this equation, the parameter $\mb$ is related to $\hbar$ as
\be
\mb^2=\frac{(m+n+1) \hbar}{2 \pi}, 
\ee
while $a$, $c$ are given by 
\be
a =\frac{m \mb}{2(m+n+1)}, \qquad c=\frac{\mb}{2(m+n+1)}. 
\ee
Since we have an explicit formula for the kernel of $\rho_{m,n}$, we can write down an explicit expression for the integral calculating the fermionic trace of this operator, which 
we will denote as $Z_{m,n}(N, \hbar)$. This expression can be put in a very convenient form if 
we use Cauchy's identity, as in \cite{kwy2,mp}, 
 \be
 \label{cauchy}
 \ba
  {\prod_{i<j}  \left[ 2 \sinh \left( {\mu_i -\mu_j \over 2}  \right)\right]
\left[ 2 \sinh \left( {\nu_i -\nu_j   \over 2} \right) \right] \over \prod_{i,j} 2 \cosh \left( {\mu_i -\nu_j \over 2} \right)}  
 & ={\rm det}_{ij} \, {1\over 2 \cosh\left( {\mu_i - \nu_j \over 2} \right)}\\
 &=\sum_{\sigma \in S_N} (-1)^{\epsilon(\sigma)} \prod_i {1\over 2 \cosh\left( {\mu_i - \nu_{\sigma(i)} \over 2} \right)}.
 \ea
  \ee
  In this way one obtains,  
  \be
  \label{zmn}
  Z_{m,n}(N, \hbar)= \frac{1}{N!}  \int_{\mathbb R^N} \frac{\rd^N p}{\mb^N} 
  \prod_{i=1}^N \left| \mypsi{a}{c}(p_i) \right|^2 \frac{\prod_{i<j} 4 \sinh \left ( \frac{\pi}{\mb}(p_i-p_j) \right )^2}{\prod_{i,j} 2 \cosh 
  \left( \frac{\pi}{\mb}(p_i-p_j)+ \im \pi C_{m,n}\right)}, 
  \ee
  where
\be
C_{m,n}=\frac{m-n+1}{2(m+n+1)}.
\ee
The above integral is real, since the kernel (\ref{rmn-ker}) is Hermitian. Although it is not manifest, it can be also checked that it 
is symmetric under the exchange $m \leftrightarrow n$. 
 
\begin{figure}
\center
\includegraphics[height=4cm]{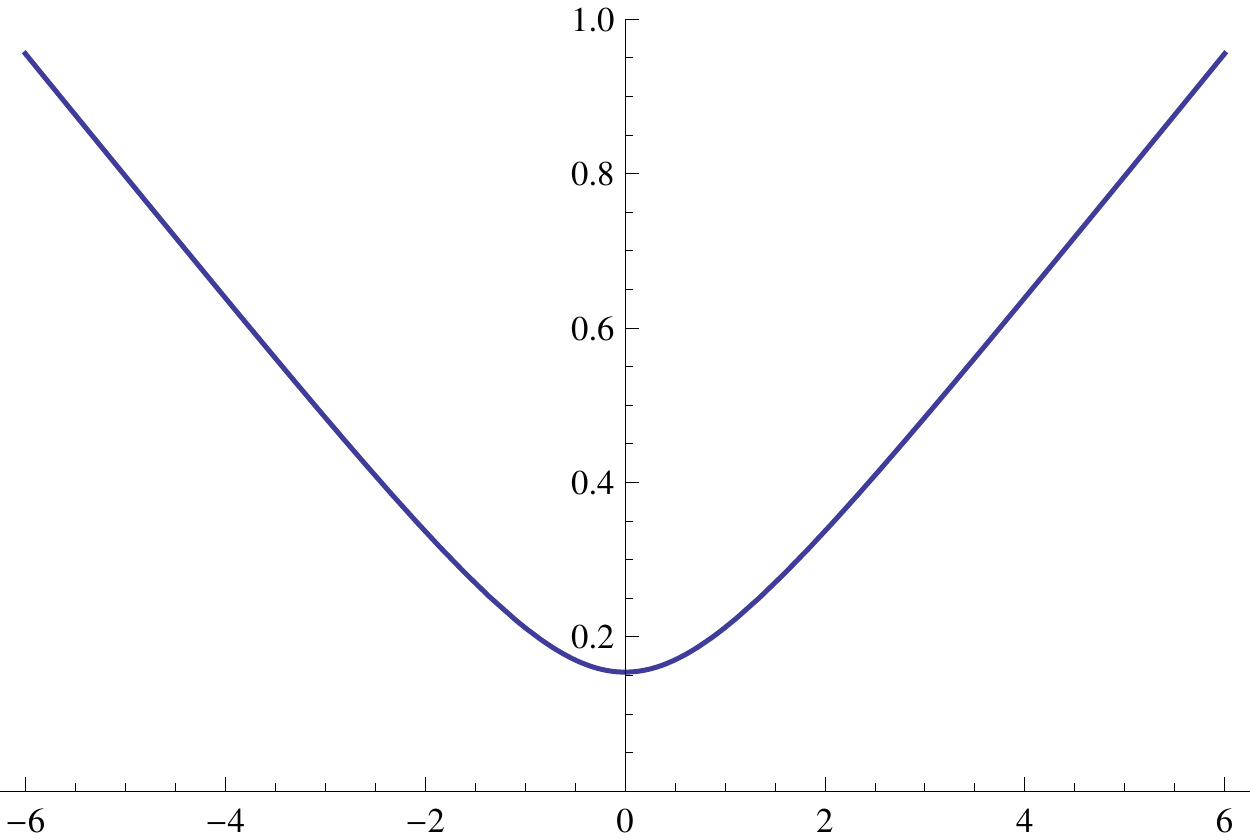}  \qquad \qquad \includegraphics[height=4cm]{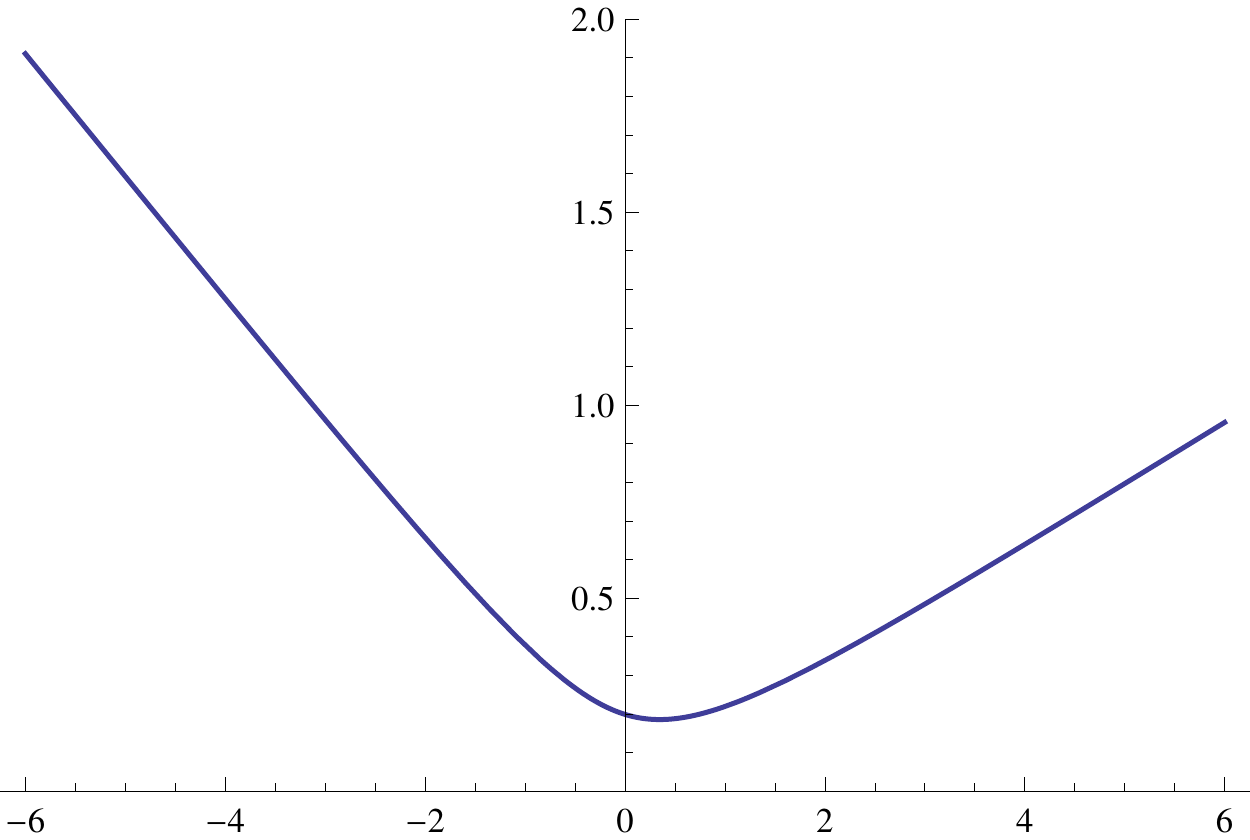}
\caption{On the left, we plot the potential (\ref{vo}) for $m=n=1$ (relevant for local $\IP^2$), while on the right we plot it for 
$m=2$, $n=1$ (relevant for local $\IF_2$).}
\label{pot-plots}
\end{figure}

We are now interested in studying the matrix integral (\ref{zmn}) in the 't Hooft limit (\ref{thooftlimit}). Therefore, 
we should understand what happens to the integrand in (\ref{zmn}) when 
$\hbar$ (or equivalently $\mb$) is large. To do this, we first change variables to
\be
\label{up}
u_i= {2 \pi \over \mb}p_i, 
\ee
and introduce the parameter
\be
\mg= {1\over \hbar}= {m+n+1\over 2 \pi} {1\over \mb^2}. 
\ee
Note that the strong coupling regime of $\hbar$ is the weak coupling regime of $\mg$. 
The crucial property to understand this regime is the self-duality of Faddeev's quantum dilogarithm, 
\be
\fad(x)=\fadm(x).
\ee
Then, we can write
\be
\ba
\left|\mypsi{a}{c}(p)\right|^2 &=   \re^{4\pi a p }{\fad(p+\im (a+c)) \over \fad(p-\im (a+c))}= \exp \left( {m u  \over 2 \pi \mg} \right) 
{ \fadm \left( \left(u +2 \pi \im (a+c)/\mb \right)/2 \pi \mb^{-1} \right) \over  \fadm \left( \left(u - 2 \pi \im (a+c)/\mb\right)/2 \pi \mb^{-1} \right) }, 
\ea
\ee
where $u$ and $p$ are related through (\ref{up}). When $\mb$ is large, $1/\mb$ is small and we can use the asymptotic expansion (see \cite{ak} for this and other properties of the 
quantum dilogarithm), 
\be\label{as-fd}
\log \fad\left( {x \over 2 \pi \mb} \right) =  \sum_{k=0}^\infty \left( 2 \pi \im \mb^2 \right)^{2k-1} {B_{2k}(1/2) \over (2k)!} {\rm Li}_{2-2k}(-\re^x), 
\ee
where $B_{2k}(z)$ is the Bernoulli polynomial. We define the {\it potential} of the matrix model as, 
\be
V_{m,n}(u, \mg)=-\mg \log |  \Psi_{a,c}(p) |^2, 
\ee
where $u$ and $p$ are related as in (\ref{up}). By using (\ref{as-fd}), we deduce that this potential has an asymptotic expansion at small $\mg$, of the form 
\be
\label{vug}
V_{m,n}(u, \mg)= \sum_{\ell \ge 0} \mg^{2\ell} V_{m,n}^{(\ell)} (u). 
\ee
The leading contribution as $\mg\rightarrow 0$ is given by the ``classical" potential,  
\be
\label{vo}
V_{m,n}^{(0)}(u)=-{m \over 2 \pi } u  - {m+n+1 \over  2\pi^2} {\rm Im} \left( {\rm Li}_2 \left(-\re^{u + \pi \im\frac{ m+1}{m+n+1}} \right) \right).
\ee
By using the asymptotics of the dilogarithm, 
\be
{\rm Li}_2(-\re^x)\approx \begin{cases} -{x^2/2}, & x\rightarrow \infty, \\
-\re^{x}, & x\rightarrow -\infty, \end{cases}
\ee
we find that
\be
V_{m,n}^{(0)}(u)\approx \begin{cases} {u \over 2 \pi}, & u\rightarrow \infty, \\
-{m  \over 2 \pi} u , & u\rightarrow -\infty. 
\end{cases}
\ee
Therefore, this is a linearly confining potential at infinity, similar to the potentials appearing in 
matrix models for Chern--Simons--matter theories \cite{mp,grassi-marino}. The classical potentials for the 
cases $m=n=1$ (relevant for local $\IP^2$) and for $m=2$, $n=1$ (relevant for local $\IF_2$) are shown in \figref{pot-plots}. 
We can now write the matrix integral as
\be
\label{zmn-bis}
Z_{m,n}(N,\hbar)=\frac{1}{N!}  \int_{\IR^N}  { \rd^N u \over (2 \pi)^N}  \prod_{i=1}^N \re^{-{1\over \mg} V_{m,n}(u_i, \mg)}  \frac{\prod_{i<j} 4 \sinh \left( {u_i-u_j \over 2} \right)^2}{\prod_{i,j} 2 \cosh \left( {u_i -u_j \over 2} + \ri \pi C_{m,n} \right)}.
\ee
In the regime in which $\hbar$ is large (in particular, in the 't Hooft limit), we can use the asymptotic expansion (\ref{vug}) of the potential to study this matrix integral. 
The expression (\ref{zmn-bis}), computing the fermionic traces of the operators $\rho_{m,n}$, is very similar to matrix models that 
have been studied before in the literature. The interaction between eigenvalues is identical to the one appearing in the generalized $O(2)$ models 
appearing in \cite{kostov-exact}, and in some matrix models 
for Chern--Simons--matter theories studied in for example \cite{kwy2}. The parameter $\mg$ corresponds to the string coupling constant, 
and the potential depends itself on $\mg$. Note however that, in the planar limit, 
only the classical part of the potential (\ref{vo}) contributes\footnote{The potentials appearing in some of the matrix models 
considered in \cite{eynard,ks,sul,ekm1,ekm2} also involve (quantum) dilogarithms, but this does not necessarily indicate a deep connection 
between these matrix models and ours, due to the reasons listed in the introduction.}.   

\subsection{Perturbative expansion}

The matrix model (\ref{zmn-bis}) admits a standard 't Hooft expansion, of the form 
\be
\label{thooftmn}
F_{m,n}(N, \hbar)=\log Z_{m,n}(N, \hbar) =\sum_{g\ge 0} \hbar^{2-2g} \CF_g^{(m,n)} (\lambda), \qquad \lambda= {N \over \hbar}. 
\ee
This can be easily seen by noting that, if we just keep the classical part of the potential, we have a generalized $O(2)$ matrix model, of the 
type considered in \cite{kostov-exact}. The corrections to 
the potential involve even powers of $\mg$, therefore they lead to corrections which preserve the form of (\ref{thooftmn}). 

We would like to compute the genus $g$ free energies $\CF_g^{(m,n)} (\lambda)$. Ideally, one would like to obtain them in closed form, as functions of the 
't Hooft parameter $\lambda$, and this might feasible via a suitable generalization of the techniques of \cite{kostov-exact}. In this paper we are interested in testing 
whether the above free energies reproduce the genus expansion of the topological string, and we will perform a more pedestrian calculation, 
by doing perturbation theory in $\mg$ to 
the very first orders. This means that we regard (\ref{zmn-bis}) as a Gaussian Hermitian matrix model, perturbed by single and double trace operators. 
The resulting perturbative expansion can then be converted, by standard means, into a weak coupling expansion around $\lambda=0$ 
of the very first $\CF_g^{(m,n)}(\lambda)$. This is very similar to the calculations done in \cite{akmv} for the lens space matrix model. 

In order to work out the expansion of the matrix model, we first have to expand the ``classical" potential $V_{m,n}^{(0)}(u)$ around its minimum. 
Let us introduce the parameter
\be
q= \exp\left( {\pi \ri \over m+n+1} \right). 
\ee
Then, the minimum of the classical potential occurs at 
\be
u_\star=\log \chi_m(q), 
\ee
where
\be
 \chi_m(q)=\frac{q^m-q^{-m}}{q-q^{-1}}.
 \ee
The value of the potential at the minimum is given by 
\be
\label{genvm}
V_{m,n}^{(0)}(u_\star)=-\frac{m}{2 \pi}\log \chi_m -\frac{m+n+1}{2\pi^2} {\rm Im}\, {\rm Li}_2 \left( -q^{m+1} \chi_m \right). 
\ee
This can be written in a more compact form by using the Bloch--Wigner function 
\be
D(z)={\rm Im \, Li_2}(z)+{\rm arg}(1-z)\log|z|, 
\ee
where arg denotes the branch of the argument between $-\pi$ and $\pi$. One finds, 
\be
V_{m,n}^{(0)}(u_\star)=-\frac{m+n+1}{2\pi^2} D(-q^{m+1}\chi_m). 
\ee
One can use the properties of the Bloch--Wigner function 
\be
D(\overline z)= -D(z), \qquad D(1-z)=-D(z), 
\ee
to verify that (\ref{genvm}) is symmetric under the exchange of $m$ and $n$. 
For example, we have
\be
V_{1,1}^{(0)}\left(u_\star=0\right)= {3 V\over 4 \pi^2}, 
\ee
where
\be
 \label{eight}
 V= 2 {\rm Im}\left( {\rm Li}_2 \left(\re^{\pi \ri \over 3} \right) \right)
 \ee
 is related to the volume of the figure-eight knot. Similarly, we find 
 \be
 V_{2,1}^{(0)}\left(u_\star={1\over2} \log(2)\right)= {2 C\over \pi^2},
 \ee
 where $C=0.915966...$ is the Catalan number.  As we will see, in order for the conjecture of \cite{ghm} to be true, (\ref{genvm}) 
 has to be related in a precise way to a natural constant 
appearing in special geometry, namely the value of the large radius K\"ahler parameter at the conifold point. 

We can now expand the full potential appearing in (\ref{zmn-bis}) around $u=u_\star$, and write the interaction term as a ``deformed" Vandermonde term, 
\be
 \frac{\prod_{i<j} 4 \sinh \left( {u_i-u_j \over 2} \right)^2}{\prod_{i,j} 2 \cosh \left( {u_i -u_j \over 2} + \ri \pi C_{m,n} \right)}= {1\over 4 I_{m,n}^2} \Delta^2(u_i) \left(1+ \CO \left(\Delta^2(u_i)\right)\right), 
 \ee
 where 
 \be
 \Delta^2(u_i)= \prod_{i<j} (u_i-u_j)^2
 \ee
 is the usual squared Vandermonde, and 
 \be
 \label{imn}
 I_{m,n}=\cos(\pi C_{m,n})=\sin \left(  \frac{\pi n}{m+1+n} \right).
 \ee
At leading order, we find a Hermitian Gaussian matrix model, and by including the corrections coming from the deformed Vandermonde and the 
potential (\ref{vug}), we can compute systematically the 't Hooft expansion of $Z_{m,n}$ around $\lambda=0$. 
By using the standard formula for the partition function of the Gaussian matrix model, 
\be
Z_{\rm G}(N,g_s)=\frac{1}{N!}\int \frac{\rd^N y}{(2 \pi)^N}\re^{-\frac{1}{2 g_s}\sum_i y_i^2} \prod_{i<j} (y_i-y_j)^2=\frac{g_s^{N^2/2}}{(2\pi)^{N/2}}G(N+1), 
\ee
where $G(z)$ is Barnes' function, as well as its asymptotic expansion at large $N$, 
we obtain the following results for the expansion of the genus $g$ free energies $\CF_g^{(m,n)}(\lambda)$ 
appearing in (\ref{thooftmn}). For the planar free energy, we find
\be
\label{planar-f}
\CF^{(m,n)}_0(\lambda)=\frac{\lambda^2}{2} \left( \log \frac{ \lambda  \pi^3}{(m+n+1)I_{m,n}^2 A_2}-\frac{3}{2}\right)-V^{(0)}_{m,n}(u_\star) \lambda +\sum_{k=3}^\infty 
f_{0,k} \lambda^k. 
\ee
In this equation, $I_{m,n}$ is given in (\ref{imn}), $A_2$ is given by 
\be
A_2=2 \pi \frac{\sin \left( \frac{\pi m}{m+n+1}\right)\sin \left( \frac{\pi}{m+n+1}\right)}{\sin \left( \frac{\pi n}{m+n+1}\right)}, 
\ee
and the values of the coefficients $f_{0,k}$ can be calculated explicitly as functions of $m,n$. The results for the very first $k$ can be found 
in Appendix \ref{app-fs}. Similarly, one finds
\be
\label{cone-sing}
\ba
\CF^{(m,n)}_1(\lambda)&=-\frac{1}{12}\log (\lambda \hbar)+\zeta'(-1)+\sum_{k=1}^\infty f_{1,k} \lambda^k, \\ 
\CF^{(m,n)}_g(\lambda)&=\frac{B_{2g}}{2g(2g-2)}\lambda^{2-2g}+\sum_{k=1}^\infty f_{g,k} \lambda^k, \qquad  g \ge 2. 
\ea
\ee
The values of $f_{g,k}$ for the very first $g$, $k$, for general $m,n$, can be also found in Appendix \ref{app-fs}. We now list some results in the case of $m=n=1$, relevant 
as we will see for local $\IP^2$. We find, 
\be
\label{exp-p2}
\ba
\CF_0^{(1,1)}(\lambda)&=\frac{\lambda^2}{2} \left( \log \left(\frac{4 \pi^2 \lambda}{9 \sqrt{3}}\right)-{3\over 2} \right)-\frac{3 V}{4 \pi^2}\lambda  -\frac{\pi^2}{9 \sqrt{3}}\lambda^3+\frac{\pi^4}{486}\lambda^4+\frac{56\pi^6}{10 935 \sqrt{3}}\lambda^5\\
&-\frac{1058 \pi^8}{492075}\lambda^6+\CO(\lambda^7), \\
\CF_1^{(1,1)}(\lambda)&=-\frac{1}{12}\log (\lambda \hbar)+\zeta'(-1)+ \frac{5 \pi^2}{18\sqrt{3}}\lambda-\frac{\pi^4}{486}\lambda^2-\frac{40 \pi^6}{2187 \sqrt{3}}\lambda^3\\
&+\frac{283 \pi^8}{32805}\lambda^4 +\CO(\lambda^5), \\
\CF_2^{(1,1)}(\lambda)&=-{1\over 240} \lambda^{-2}+\frac{4 \pi^6}{405\sqrt{3}}\lambda-\frac{3187 \pi^8}{492075}\lambda^2+\CO(\lambda^3). 
\ea
\ee
For $m=2$, $n=1$, which is relevant for local $\IF_2$, we obtain, 
\be
\label{exp-f2}
\ba
\CF_0^{(2,1)}(\lambda)&=\frac{\lambda^2}{2} \left( \log \left(\frac{\pi^2 \lambda}{4} \right)-{3\over 2} \right)-\frac{2 C}{\pi^2}\lambda  -\frac{\pi^2}{12}\lambda^3+\frac{5 \pi^4}{288}\lambda^4-\frac{7 \pi^6}{960}\lambda^5 
+\frac{733 \pi^8}{172800}\lambda^6+\CO(\lambda^7), \\
\CF_1^{(2,1)}(\lambda)&=-\frac{1}{12}\log (\lambda \hbar)+\zeta'(-1)+ \frac{\pi^2}{6}\lambda-\frac{5\pi^4}{288}\lambda^2+\frac{\pi^6}{576}\lambda^3 
+\frac{53 \pi^8}{17280}\lambda^4+ \CO(\lambda^5), \\
\CF_2^{(2,1)}(\lambda)&=-{1\over 240} \lambda^{-2}+\frac{\pi^6}{240}\lambda-\frac{421 \pi^8}{57600}\lambda^2+\CO(\lambda^3).
\ea
\ee 

\sectiono{Testing the matrix model}

\subsection{What should we expect?}

The conjecture of \cite{ghm} gives a very precise prediction for the 't Hooft expansion (\ref{free-expansion}) of the matrix models arising from 
the trace class operators. To understand this prediction, we have to summarize some of the results of \cite{ghm}. According to the conjecture of \cite{ghm}, the basic 
quantity determining the spectral properties of the operator $\rho_S$ is the modified grand potential $J_S(\mu, \xi_i, \hbar)$. This grand potential depends 
on the ``fugacity" $\mu$, which is related to the variable $\kappa$ entering in (\ref{f-det}) as 
\be
\kappa=\re^\mu, 
\ee
as well as on the parameters $\xi_i$ appearing in the operator $\mathsf{O}_S$, which we will collect in a vector $\boldsymbol{\xi}$. The modified grand potential is 
determined by the enumerative geometry of the Calabi--Yau $X$. 
We first need a dictionary between the parameters $\mu$, $\boldsymbol{\xi}$, and the parameters appearing in the enumerative geometry of $X$. 
First of all, we remember that the mirror curve (\ref{ex-W}) involves 
a modulus $\tilde u$. Typically, in mirror symmetry one uses the related modulus
\be
\label{z-mod}
z={1\over \tilde u^r}, 
\ee
where, the value of $r$ is determined by the geometry of $X$. There is a corresponding flat coordinate $T$, determined by 
\be
\label{Tmm}
-T=\log z+ \CO(z). 
\ee
Let us now consider the K\"ahler parameters of a general local del Pezzo, $T_i$, in an arbitrary basis. 
They can be always be written as a linear combination of the K\"ahler parameter $T$ corresponding 
to $z$, and the K\"ahler parameters $T_{\xi_i}$, which are algebraic functions of the parameters $\xi_i$ appearing 
in the mirror curve. We have then, 
\be
\label{genkal}
T_i = {c_i \over r} T + \alpha_{ij} T_{\xi_j}. 
\ee
The dictionary is now given as follows. On top of the standard mirror map (\ref{Tmm}), there is a quantum mirror map 
\cite{acdkv} of the form 
\be
\label{qmmap}
-T(\hbar)=\log(z)+ \sum_{\ell \ge 1} \widehat a_\ell (\hbar) z^\ell. 
\ee
The ``effective" $\mu$ parameter is defined by an expansion similar to this one, 
\be
\label{mueff}
\mu_{\rm eff}= \mu- {1\over r} \sum_{\ell \ge 1} (-1)^{r \ell} \widehat a_\ell (\hbar) \re^{-r \ell \mu}. 
\ee
The K\"ahler parameters $T_i$ are then related to the parameters $\mu$, $\xi_j$ by the following equation, 
\be
T_i = c_i \mu_{\rm eff}+ \alpha_{ij} T_{\xi_j}. 
\ee
We can now write down the general expression for the modified grand potential. It is of the form, 
 \be
 \label{jx-masses}
 J_S (\mu, \boldsymbol{\xi}, \hbar) = J^{({\rm p})} (\mu_{\rm eff}, \xi_i, \hbar) + J_{\rm M2} (\mu_{\rm eff}, \boldsymbol{\xi}, \hbar) + J_{\rm WS}(\mu_{\rm eff}, \boldsymbol{\xi}, \hbar). 
 \ee
Here, $J^{({\rm p})} (\mu, \boldsymbol{\xi}, \hbar) $ is the perturbative part, which is a cubic polynomial in $\mu$:
 \be
 \label{jp}
 J^{({\rm p})} (\mu, \boldsymbol{\xi}, \hbar)={C \over 6 \pi \hbar } \mu^3 + {D(\boldsymbol{\xi}) \over \hbar} \mu^2 
 +\left( {B_0(\boldsymbol{\xi}) \over \hbar}+ B_1(\boldsymbol{\xi}) \hbar \right) \mu + A(\boldsymbol{\xi}, \hbar). 
 \ee
 The constants appearing in this expression can be obtained from a semiclassical analysis of the operator $\rho_S$ (see \cite{ghm} for details). 
 The function $A(\boldsymbol{\xi}, \hbar)$ is not known in closed form, but in some examples there are educated guesses for it. The ``membrane" part of the potential has the form, 
 \be
 \label{membrane}
 J_{\rm M2}(\mu_{\rm eff}, \boldsymbol{\xi}, \hbar)= \mu_{\rm eff}    \widetilde{J}_b(\mu_{\rm eff},\boldsymbol{\xi}, \hbar)+\widetilde{J}_c(\mu_{\rm eff},\boldsymbol{\xi}, \hbar), 
 \ee
where
 \be
 \ba
 \widetilde{J}_b(\mu_{\rm eff},\boldsymbol{\xi}, \hbar)&=-{1\over 2 \pi} \sum_{j_L, j_R} \sum_{w, {\bf d} }\left(  {\bf c} \cdot {\bf d} \right) N^{{\bf d}}_{j_L, j_R} 
 \frac{\sin\frac{\hbar w}{2}(2j_L+1)\sin\frac{\hbar w}{2}(2j_R+1)}{2 w \sin^3\frac{\hbar w}{2}} \re^{-w {\bf d}\cdot{\bf  T}}, \\
  \widetilde{J}_c(\mu_{\rm eff},\boldsymbol{\xi}, \hbar)&=-{1\over 2 \pi} \sum_{i,j} \sum_{j_L, j_R} \sum_{w, {\bf d} } d_i \alpha_{ij} T_{\xi_j} N^{{\bf d}}_{j_L, j_R} 
 \frac{\sin\frac{\hbar w}{2}(2j_L+1)\sin\frac{\hbar w}{2}(2j_R+1)}{2 w \sin^3\frac{\hbar w}{2}} \re^{-w {\bf d}\cdot{\bf  T}}\\
 &+{1\over 2 \pi} \sum_{j_L, j_R} \sum_{w, {\bf d} } \hbar^2 {\partial \over \partial \hbar} \left[  \frac{\sin\frac{\hbar w}{2}(2j_L+1)\sin\frac{\hbar w}{2}(2j_R+1)}{2 \hbar w^2 \sin^3\frac{\hbar w}{2}} \right] N^{{\bf d}}_{j_L, j_R} \re^{-w {\bf d}\cdot{\bf  T}}. 
 \ea
 \ee
 In this equation, $N^{{\bf d}}_{j_L, j_R}$ are the refined BPS invariants of the CY $X$, ${\bf T}= (T_1, T_2, \cdots)$ 
 is the vector of K\"ahler parameters, ${\bf c}$ is the vector 
 of constants $c_i$ appearing in (\ref{genkal}), and ${\bf d}$ is the vector of degrees. Finally, the worldsheet part of the 
 modified grand potential is 
\be
\label{wsj}
J_{\rm WS}(\mu_{\rm eff}, \boldsymbol{\xi}, \hbar)= \sum_{g\ge 0} \sum_{{\bf d}, v} n^{\bf d}_g {1\over v} \left( 2 \sin {2 \pi^2 v\over \hbar} \right)^{2g-2} (-1)^{{\bf d} \cdot {\bf B}} 
\re^{- {2 \pi \over \hbar } v {\bf d} \cdot {\bf T}}.
\ee
In this equation, $n^{\bf d}_g$ are the Gopakumar--Vafa invariants of $X$, while ${\bf B}$ is a $B$-field, which is related to the anticanonical class of $S$, and necessary for 
the cancellation of poles in $J_S(\mu, \boldsymbol{\xi}, \hbar)$ \cite{hmmo}.

\begin{figure}
\center
\includegraphics[height=6cm]{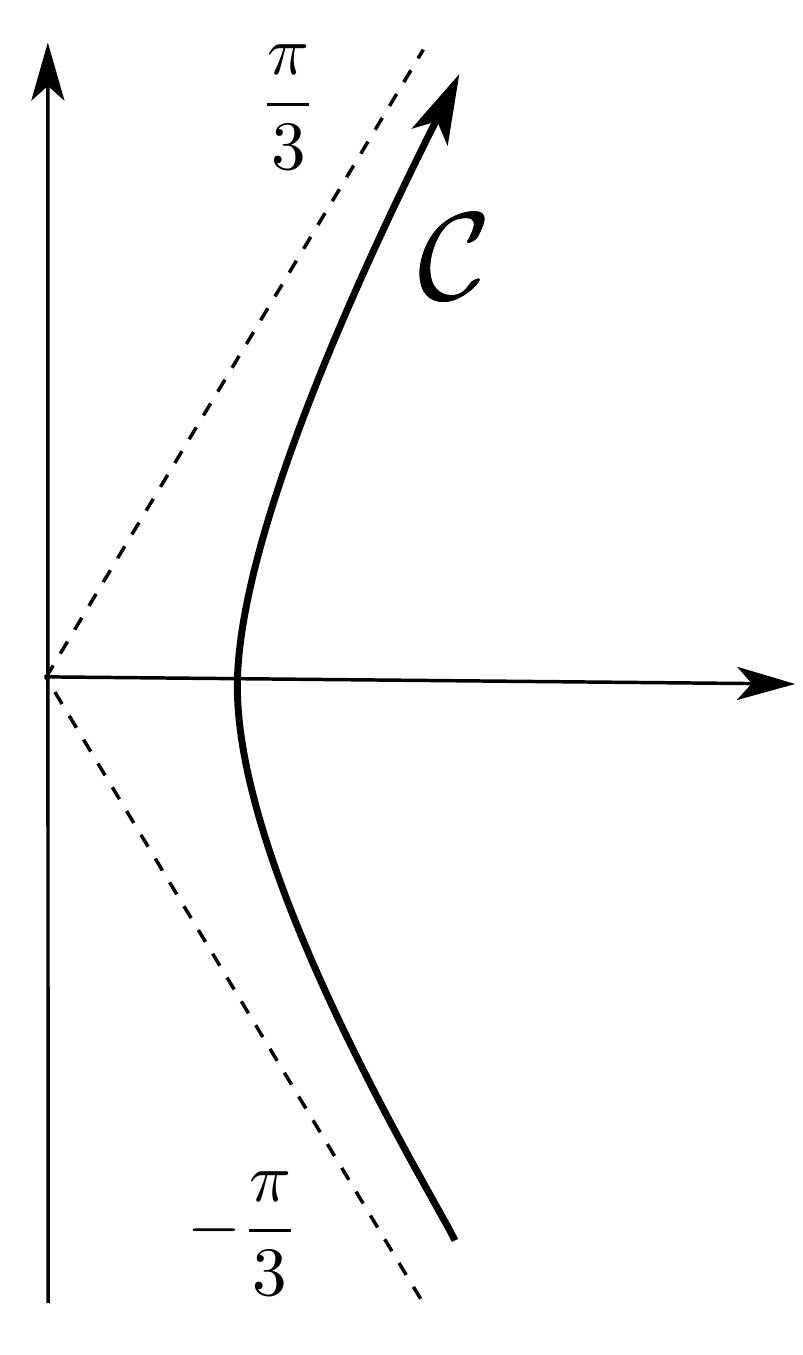}  
\caption{The contour $\CC$ in the complex $\mu$ plane, which can be used to calculate the fermionic trace $Z(N, \hbar)$ from the modified grand potential.}
\label{airy-c}
\end{figure}

According to the conjecture of \cite{ghm}, the spectral determinant of the operator $\rho_S$ is given by 
\be
\label{xi-j}
\Xi_S(\mu, \hbar)= \sum_{n \in \IZ} \re^{J_S(\mu+2 \pi \ri n, \boldsymbol{\xi}, \hbar)}. 
\ee
As first shown in \cite{hmo}, in the context of ABJM theory, this representation leads to a very convenient formula for the 
fermionic trace $Z_S(N, \hbar)$ as an integral transform of the modified grand potential,
\be
\label{zn-airy}
Z_S(N, \hbar)={1\over 2 \pi \ri} \int_\CC \re^{J_S(\mu,\boldsymbol{\xi}, \hbar) - N \mu} \rd \mu. 
\ee
The contour $\CC$ appearing in this integral is shown in 
\figref{airy-c}, and in view of the cubic behavior of $J_S(\mu, \boldsymbol{\xi}, \hbar)$, it leads to a convergent integral (this is the contour used to define the 
Airy function). 

As already pointed out in \cite{ghm}, the above conjectural results lead to a very precise formula for the 't Hooft expansion of $Z_S(N, \hbar)$. 
To obtain this expansion, we follow a procedure similar to what was done in \cite{kkn}. 
We first note that the function $J_S(\mu, \boldsymbol{\xi}, \hbar)$ has itself a 't Hooft limit in which 
\be
\label{j-limit}
\mu \rightarrow \infty, \qquad \hbar \rightarrow \infty, \qquad {\mu \over \hbar}= \zeta \, \, \, {\rm fixed}. 
\ee
It is clear that this double scaling is needed if we want the integral in the r.h.s. of (\ref{zn-airy}) to be non-trivial as 
$\hbar \rightarrow \infty$. What is the effect 
of this limit on $J_S(\mu, \boldsymbol{\xi}, \hbar)$? First of all, we have that 
\be
{\mu_{\rm eff} \over \hbar} \rightarrow \zeta, 
\ee
and all the exponential corrections to (\ref{mueff}) {\it vanish}. Similarly, the membrane corrections in (\ref{membrane}) 
vanish as well, since they are exponentially small 
as $\mu \rightarrow \infty$. The only surviving terms come from the perturbative and the worldsheet parts of the modified grand potential. 
As we will see in examples, the function $A(\boldsymbol{\xi}, \hbar)$ has an expansion as $\hbar \rightarrow \infty$ of the form 
\be
A(\boldsymbol{\xi}, \hbar)= \sum_{g\ge 0} A_g(\boldsymbol{\xi})\hbar^{2-2g}, 
\ee
where $A_1(\boldsymbol{\xi})$ includes as well a logarithmic dependence on $\hbar$. We conclude that, in the 't Hooft limit (\ref{j-limit}), the modified grand potential has the expansion, 
\be
J^{\text{'t Hooft}}_S(\zeta, \boldsymbol{\xi}, \hbar) = \sum_{g=0}^\infty J_g^S(\zeta, \boldsymbol{\xi}) \hbar^{2-2g}, 
\ee
where
\be
\label{jg-par}
\ba
J_0^S(\zeta, \boldsymbol{\xi})&={C\over 6 \pi } \zeta^3 + B_1 \zeta +A_0(\boldsymbol{\xi})+ {D(\boldsymbol{\xi}) \over \hbar} +{1\over 16 \pi^4} F^{\rm inst}_0\left( t, {2 \pi T_{\xi_i} \over \hbar} \right), \\
J_1^S(\zeta, \boldsymbol{\xi})&=B_0 \zeta +  A_1(\boldsymbol{\xi})+F_1^{\rm inst}\left( t, {2 \pi T_{\xi_i} \over \hbar} \right), \\
J_g^S(\zeta, \boldsymbol{\xi})&= A_g(\boldsymbol{\xi}) + (4 \pi^2)^{2g-2} F_g^{\rm inst}\left( t, {2 \pi  T_{\xi_i}  \over \hbar} \right). 
\ea
\ee
Here, we have introduced the variable
\be
\label{tr}
t= 2 \pi r \zeta, 
\ee
and $F^{\rm inst}_g(t, t_{\xi_i})$ is the standard genus $g$ topological string free energy, expressed 
in terms of the conventional K\"ahler parameters $t$, $t_{\xi_i}$ (these correspond to the parameters $T$, $T_{\xi_i}$, up to a rescaling by $2\pi/\hbar$). 
We now note that, in order to perform the 't Hooft limit, we must also make a choice for the scaling of the parameters $T_{\xi_i}$. If these scale
with $\hbar$ when $\hbar\rightarrow \infty$, the 't Hooft limit of $J_g^S(\zeta, \xi)$ reproduces the standard topological string free energy at genus $g$. However, 
we might also want to keep the $T_{\xi_i}$ fixed. In that case, in order to obtain the large $\hbar$ expansion, we have to re-expand the 
functions $J_g^S(\zeta, \xi)$, and one finds 
\be
J^{\text{'t Hooft}}_S(\zeta, \boldsymbol{\xi}, \hbar) = \sum_{g=0}^\infty \tilde J_g^S(\zeta, \boldsymbol{\xi}) \hbar^{2-2g}, 
\ee
where, for the very first orders, 
\be
\ba
\tilde J_0^S(\zeta, \boldsymbol{\xi})&={C\over 6 \pi } \zeta^3 + B_1 \zeta +A_0(\boldsymbol{\xi})+ {D(\boldsymbol{\xi}) \over \hbar} +{1\over 16 \pi^4} F^{\rm inst}_0\left( t, 0 \right), \\
\tilde J_1^S(\zeta, \boldsymbol{\xi})&=B_0 \zeta +  A_1(\boldsymbol{\xi})+F_1^{\rm inst}\left( t, 0 \right) +{1\over 8 \pi^2}T_{\xi_i} T_{\xi_j} {\partial^2 F^{\rm inst}_0\left( t, 0 \right) 
\over \partial t_{\xi_i} \partial t_{\xi_j}}. 
\label{mod-js}
\ea
\ee
Similar expressions can be of course obtained for the higher order functions $\tilde J_g^S(\zeta, \boldsymbol{\xi})$, involving derivatives of the $F^{\rm inst}_{g'}$ with $g'<g$. 
Here we have assumed that derivatives of the free energies w.r.t. an odd number of parameters $t_{\xi_i}$, and evaluated at $t_{\xi_i}=0$, vanish, in order 
to have an expansion involving only even powers of $\hbar$. This is the case in the example of local $\IF_2$, which will be analyzed in detail later on, but might not 
be a universal fact. The formulae above can be easily modified if this is not the case, and the conjecture of \cite{ghm} would 
then imply that the 't Hooft expansion of the fermionic traces has odd powers of $\hbar$. This does not happen for the operators $\rho_{m,n}$ that we are analyzing 
in this paper, but might happen in other situations. 

In any case, the formulae of \cite{ghm} lead to precise expressions for the 't Hooft limit of the modified grand potential. One interesting 
aspect of this limit is that it only keeps the worldsheet instanton corrections, which are precisely the most difficult ingredients from the point 
of view of spectral theory. In order to obtain the 't Hooft expansion of the fermionic trace, 
\be
\log Z_S(N, \hbar)= \sum_{g \ge 0} \CF_g^S(\lambda, \boldsymbol{\xi}) \hbar^{2-2g}, 
\ee
we still have to calculate the integral in (\ref{zn-airy}). 
It can be evaluated in a systematic asymptotic expansion at large $\hbar$ by using the saddle-point 
approximation. At leading order, we find that the 't Hooft parameter is given by 
\be
\label{l-jo}
\lambda= {\partial J^S_0( \zeta, \boldsymbol{\xi}) \over \partial \zeta}. 
\ee
This determines $\lambda$ as a function of $\zeta$, and conversely, $\zeta$ as a function of $\lambda$. The genus zero 
free energy $\CF^S_0( \lambda, \boldsymbol{\xi})$ is then given, as usual, by a Legendre transform, 
\be
\CF^S_0(\lambda, \boldsymbol{\xi})= J^S_0(\zeta(\lambda), \boldsymbol{\xi})- \lambda \zeta(\lambda). 
\ee
We obtain in particular 
\be
\label{dual-per}
{\partial \CF^S_0\over \partial \lambda}= - \zeta. 
\ee
The next-to-leading order correction 
$\CF^S_1(\lambda, \boldsymbol{\xi})$ is given by the one-loop approximation to the integral (\ref{zn-airy}), 
\be
\label{gones}
\CF^S_1(\lambda, \boldsymbol{\xi})= J^S_1(\zeta(\lambda), \boldsymbol{\xi})- {1\over 2} \log \left( 2 \pi {\partial^2 J^S_0 \over \partial \zeta^2} \right). 
\ee
Higher corrections can be computed in a straightforward way, and we note that, when the parameters $T_{\xi_i}$ are fixed 
in the 't Hooft limit, we have to use the modified formulae (\ref{mod-js}). 

The calculation of higher corrections 
is better understood if we realize that the integral (\ref{zn-airy}) 
has precisely the form appropriate to implement a symplectic transformation of the topological string free energies, 
as explained in \cite{abk}\footnote{Note however that the integral transform appearing in \cite{abk} is purely formal, and relates 
two asymptotic power series, while the integral (\ref{zn-airy}) has a precise non-perturbative meaning.}. 
The equations (\ref{l-jo}) and (\ref{dual-per}) indicate that the integral (\ref{zn-airy}) is implementing an $S$ transformation, 
in which the variable $\zeta$ in $J^S_g(\zeta, \boldsymbol{\xi})$ becomes the dual variable $-\partial_\lambda \CF_g^S(\lambda, \boldsymbol{\xi})$. 
From the point of view of the topological string, the $J^S_g(\zeta, \boldsymbol{\xi})$ are free energies in the so-called large radius frame 
(where we have an interpretation in terms of a worldsheet instanton 
expansion). The $S$ transformation, in these local del Pezzo geometries, takes the large radius frame to the {\it conifold} frame. 
We conclude that, according to the conjecture of \cite{ghm}, {\it the functions $\CF^S_g(\lambda, \boldsymbol{\xi})$ appearing in the 't Hooft expansion of the 
fermionic traces (i.e. in the 't Hooft expansion of the matrix model (\ref{znmm})) are the genus $g$ free energies of the topological string in the conifold frame} (up to normalizations 
and integration constants, as we will see). 

The appearance of the conifold frame is completely natural in view of the matrix model representation. Topological string theory in the conifold 
frame is known to have a universal structure: when the free energies are expanded around the conifold point, the leading 
singularities are of the form $t_c^{2-2g}$, where $t_c$ is the flat coordinate around the conifold, and the coefficients of these singularities are determined by the 
$c=1$ string free energy \cite{gvcone}. There is in addition a ``gap" condition, \cite{hk}, which says that the corrections to 
this leading singularity involve only non-negative powers of $t_c$. It has been known for some time that this structure is precisely what is found in the 
weak 't Hooft coupling expansion of a perturbed Gaussian Hermitian matrix model. Therefore, the appearance of the conifold 
frame in the above derivation is already what one should expect from the existence of an explicit matrix model description. 

In summary, if the conjecture of \cite{ghm} is true, the matrix model (\ref{znmm}) has a 't Hooft expansion, given by the genus $g$ 
topological string free energies in the conifold frame. We will now test this prediction in detail 
in the case of the matrix integral (\ref{zmn-bis}). For $n=m=1$, this is just local $\IP^2$, while for $n=1$, $m=2$ this is 
local $\IF_2$ for a specific value of the ``mass" parameter, namely $\xi=0$. In other words, according to the conjecture of \cite{ghm}, 
we should have 
\be
\label{conj-eq}
\CF_g^{(1,1)}(\lambda)=\CF_g^{\IP^2}(\lambda), \qquad \CF_g^{(2,1)}(\lambda)=\CF_g^{\IF_2}\left(\lambda, \xi=0 \right), 
\ee
for all $g\ge 0$. The l.h.s. of these conjectured equalities has been computed, in an expansion around $\lambda=0$, in (\ref{exp-p2}) and (\ref{exp-f2}). 
In the next two sections, we will use topological string theory to compute the r.h.s. of (\ref{conj-eq}) and check to the available order that these two functions are indeed equal.

\subsection{Local $\IP^2$}

\begin{figure}
\center
\includegraphics[height=5.5cm]{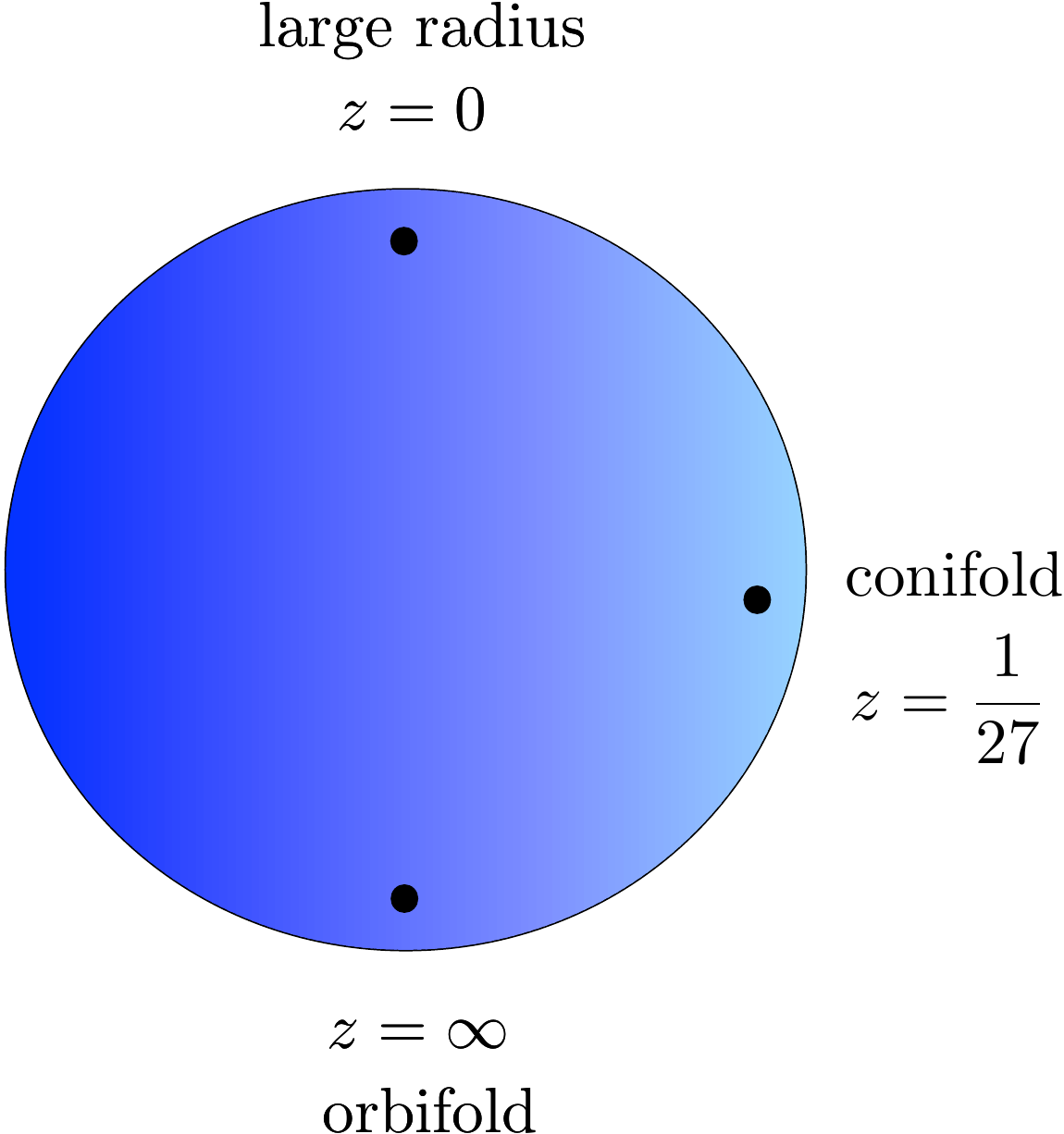}  
\caption{The moduli space of local $\IP^2$ can be parametrized by a single complex variable $z$, and it has three special points: the large radius point at $z=0$, the 
conifold point at $z=1/27$, and the orbifold point at $z=\infty$.}
\label{p2-modulus}
\end{figure}
The modified grand potential of local $\IP^2$ has been determined in \cite{ghm} in great detail. In this case, there are no mass parameters $\xi_i$. In order to write 
down the results, let us recall some elementary facts about the special geometry of local $\IP^2$, relevant for our calculation. The Picard--Fuchs equation determining 
the periods is 
\be
\left( \theta^3 -3 z (3 \theta+2) (3 \theta+1)\theta \right) \Pi=0, 
\ee
where 
\be
\theta= z{\rd \over \rd z},
\ee
and $z$ parametrizes the moduli space, which has three special points: $z=0$ is the large radius point, $z=1/27$ is the conifold point, and $z=\infty$ is the orbifold point (see \figref{p2-modulus}). 
A basis of solutions around the large radius point $z=0$ is given by
\be
\ba
\omega_1(z)&=\log(z) + 6z + 45 z^2+ 560 z^3+ \CO(z^4), \\
\omega_2(z)&=\frac{\log ^2(z)}{6}+ {\log (z) \over 3} \left( 6z + 45 z^2+ 560 z^3+ \CO(z^4) \right) 
+ 3z +\frac{141}{4}z^2+{1486\over 3} z^3+\CO(z^4). 
\ea
\ee
These periods determine the large radius, genus zero free energy $F_0(t)$ as
\be
\label{spec}
\ba
-t&= \omega_1(z), \\
{\partial F_0 \over \partial t}&= \omega_2(z), 
\ea
\ee
which gives
\be
F_0(t)= {t^3 \over 18}- 3 \re^{-t}-{45 \over 8} \re^{-2t}-{244 \over 9} \re^{-3t} +\CO(\re^{-4t}).  
\ee
Note that the signs are not the standard ones. This is due to a non-trivial $B$-field which has to be turned on in (\ref{wsj}) in 
order to obtain a consistent modified grand potential, as first noted in \cite{hmmo}. 

We can now write down the 't Hooft limit of the modified grand potential. The constants appearing in the perturbative piece (\ref{jp}) are \cite{ghm}
\be
C={9 \over 2}, \qquad D=0, \qquad B_0={\pi \over 2}, \qquad B_1=-{1\over 16 \pi}. 
\ee
On the other hand, the function $A(\hbar)$ is given by
\be
\label{ah-p2}
A(\hbar) ={3 A_{\rm c}(\hbar/\pi)- A_{\rm c}(3\hbar/\pi) \over 4}, 
\ee
where \cite{hanada} (see also \cite{ho})
 \be
\label{ak}
A_{\rm c}(k)= \frac{2\zeta(3)}{\pi^2 k}\left(1-\frac{k^3}{16}\right)
+\frac{k^2}{\pi^2} \int_0^\infty \frac{x}{\re^{k x}-1}\log(1-\re^{-2x})\rd x.
\ee
This function has a large $k$ expansion of the form, 
\be
A_{\rm c}(k)=-{k^2\over 8 \pi^2} \zeta(3)+ {1\over 2} \log(2) + 2 \zeta'(-1) +{1\over 6} \log \left({\pi \over 2k}\right) + \sum_{g\ge 2} \left({2 \pi \over k}\right)^{2g-2}  4^g (-1)^{g-1} c_g, 
\ee
where
\be
c_g={B_{2g} B_{2g-2} \over (4g) (2g-2)(2g-2)!}.
\ee
We conclude that 
\be
\label{ags}
\ba
A_0&={3 \zeta(3) \over 16 \pi^4}, \\
A_1&=-{1\over 12} \log(\hbar) + \zeta'(-1)+ {1\over 6} \log(2 \pi) + {1\over 24} \log (3), \\
A_g&= (4 \pi^2)^{2g-2} \left( 3- 3^{2-2g} \right) (-1)^{g-1} c_g, \qquad g\ge 2, 
\ea
\ee
\begin{figure}
\center
\includegraphics[height=5cm]{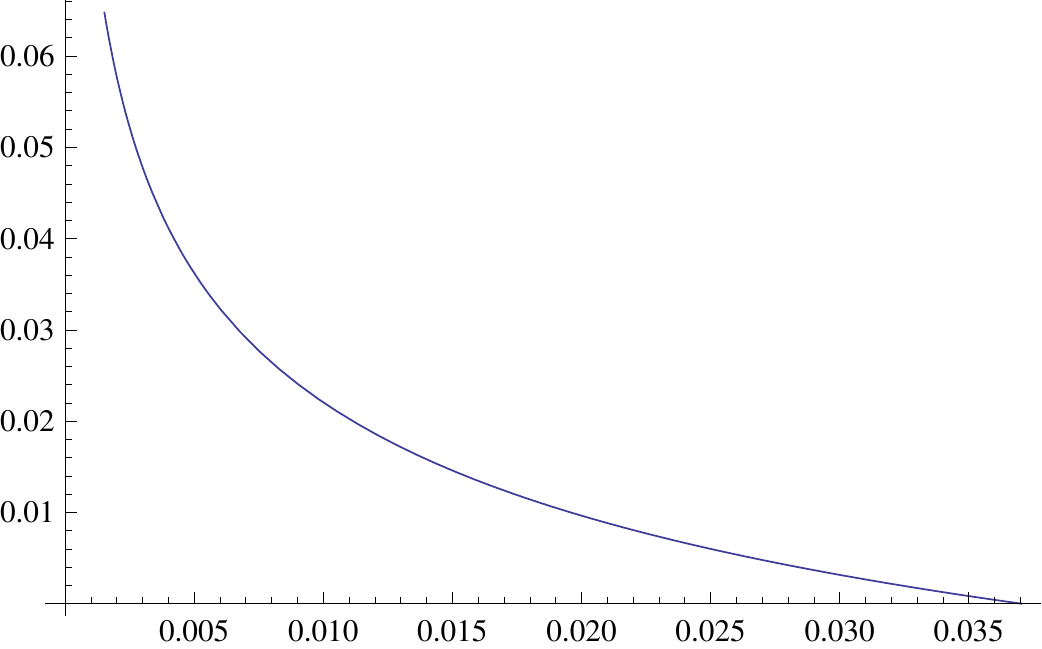} 
\caption{The 't Hooft parameter $\lambda=N/\hbar$ as a function of the modulus $0\le z\le 1/27$. The large radius point $z=0$ corresponds to 
strong 't Hooft coupling, while $z=1/27$, the conifold point, is the weakly coupled theory.}
\label{par-p2}
\end{figure}
The function $J_0^{\IP^2}(\zeta)$ is expressed more conveniently in terms of the variable $t$ appearing in (\ref{tr}), which  
is now given by, 
\be
\label{t-zeta}
t= 6 \pi \zeta,  
\ee
since $r=3$ in this geometry. We find, 
\be
J^{\IP^2}_0(\zeta)= {1\over 16 \pi^4} \left( F_0(t) +3 \zeta(3)- {\pi^2 t \over 6} \right).
\ee
The 't Hooft parameter is now given by (\ref{l-jo}), which reads in this case, 
\be
{ 8 \pi^3 \over 3} \lambda= {\partial F_0 \over \partial t} -{\pi^2 \over 6}. 
\ee
The r.h.s. of this equation is nothing but the vanishing period at the conifold point. Therefore, the 't Hooft parameter varies between $0$ and $\infty$ as $z$ varies 
between $1/27$ and $0$, as shown in \figref{par-p2}. The region around the large radius point $z=0$ in CY moduli space corresponds to 
strong 't Hooft coupling, while the region around $z=1/27$, the conifold point, corresponds to the weakly coupled theory. Since we are interested in expanding the 
free energies around $\lambda=0$, we have to analyze the theory around the conifold point. To do this, we define the variable
\be
\label{ycoord}
y=1-27 z. 
\ee
We now solve the Picard--Fuchs equation near the conifold point, i.e. near $y=0$. There are again two independent periods. One of them, which we will denote by 
$t_c(y)$, is a flat coordinate near the conifold point. It is given by the power series expansion
\be
t_c(y)=y+\frac{11 y^2}{18}+\frac{109 y^3}{243}+\frac{9389 y^4}{26244}+\frac{88351 y^5}{295245}+\frac{823187 y^6}{3188646}+\CO\left(y^7\right), 
\ee
and is related to the 't Hooft parameter as 
\be
\lambda= {{\sqrt{3}} \over 12 \pi^2} t_c(y). 
\ee
Therefore, as announced above, the 't Hooft parameter defined by the fermionic traces is a flat coordinate at the conifold point, proportional to $t_c(y)$.  

The period $\omega_1(z)$ defines the genus zero free energy $\CF^{\IP^2}_0(\lambda)$. Indeed, from (\ref{dual-per}), (\ref{t-zeta}) and (\ref{spec}), we find
\be
\label{der-om}
{\partial \CF^{\IP^2}_0(\lambda) \over \partial \lambda}= {\omega_1(z) \over 6 \pi}. 
\ee
This function is, up to normalizations and integration constants, the genus zero free energy of local $\IP^2$ in the conifold frame, and it has 
been computed in for example \cite{hkr}. In order to expand it around $\lambda=0$, we need the expansion of the period $\omega_1(z)$ around the conifold point. 
This is a standard exercise in special geometry and one finds, 
\be
\label{omega1}
\omega_1(z)= -c+{ {\sqrt 3}\over 2 \pi} \biggl( t_c(y) \log\left( { y \over 3 \log(3) +1}\right)+ s\left(y \right) \biggr), 
\ee
where
\be
s(y)=\frac{7 y^2}{12}+\frac{877 y^3}{1458}+\frac{176015 y^4}{314928}+\frac{9065753 y^5}{17714700}+\frac{17960917 y^6}{38263752}+\CO\left(y^7\right), 
\ee
and the constant $c$ is given by 
\be
\label{ct}
\ba
c&=-\frac{2}{9} \, _4F_3\left(1,1,\frac{4}{3},\frac{5}{3};2,2,2;1\right)+3 \log (3)={\rm Re}\left\{ \frac{{\sqrt{3}}}{2 \pi} 
G_{3,3}^{2,2}\left(-1\left|
\begin{array}{c}
 \frac{1}{3},\frac{2}{3},1 \\
 0,0,0
\end{array}
\right.\right)  \right\}\\
& \approx 2.90759
\ea
\ee
The relationship (\ref{der-om}) can be integrated to obtain the genus zero free energy, up to an integration constant. This constant 
can be determined as follows. The value of $\CF_0(\lambda)$ at $\lambda=0$ is given by
\be
\CF^{\IP^2}_0(0)=J^{\IP^2}_0(\zeta(0))= {1\over 16 \pi^4} \left( F_0(c) -{\pi^2 c \over 6} + 3\zeta(3)\right), 
\ee
where we used that 
\be
 t\left(z={1\over 27}\right)=c. 
\ee
The value of $J^{\IP^2}_0(\zeta(0))$ can be obtained numerically with high precision and we find that $\CF^{\IP^2}_0(0)$ {\it vanishes}, i.e. we find that
\be
F_0(c) -{\pi^2 c \over 6}=-3 \zeta(3). 
\ee
When all this is taken into account, we find the expansion,  
\be
\label{final-f0}
\ba
\CF^{\IP^2}_0(\lambda)&=-{c\over 6 \pi} \lambda+ \lambda ^2 \left(\frac{\log (\lambda )}{2}-\frac{3}{4}+\log \left(\frac{2 \pi }{3 \sqrt[4]{3}}\right)\right)-\frac{\pi ^2 \lambda ^3}{9 \sqrt{3}}+\frac{\pi ^4
   \lambda ^4}{486}+\frac{56 \pi ^6 \lambda ^5}{10935 \sqrt{3}}\\
  & -\frac{1058 \pi ^8 \lambda ^6}{492075}+\frac{3392 \pi ^{10} \lambda ^7}{2066715 \sqrt{3}}-\frac{208744 \pi ^{12} \lambda ^8}{1171827405}+\CO\left(\lambda ^9\right). 
   \ea
   \ee
This expansion should {\it exactly} agree with the expansion of $\CF_0^{(1,1)}(\lambda)$ calculated in (\ref{exp-p2}), and it does, 
provided the number $c$ given in (\ref{ct}) satisfies 
\be
c={9 V \over 2 \pi}. 
\ee
Amazingly, this is a true identity \cite{rv,dk}! (see also \cite{yang}). This encapsulates the power of the conjecture of \cite{ghm}: the constant $c$ 
comes from the world of topological string theory, and it gives the value of the K\"ahler parameter $t$ at the conifold point. 
The constant $V$ comes from the world of trace class operators arising from mirror curves, 
which lead to quantum dilogarithms by the results of \cite{kas-mar}. 
For the conjecture of \cite{ghm} to work, the two numbers coming from these two different worlds have to agree. And they do.

The agreement between $\CF^{\IP^2}_0(\lambda)$ and $\CF^{(1,1)}_0(\lambda)$ (at the order we have been able to check) is 
a non-trivial analytic test of the conjecture put forward in \cite{ghm}. Specifically, this calculation tests one of the most surprising 
claims of \cite{ghm}, namely that the non-perturbative corrections appearing in the spectral 
theory of the operator $\rho_{\IP^2}$ are encoded in the {\it standard} topological string. 

Let us now consider the genus one free energy. By using well-known results for the B-model of local $\IP^2$, one finds 
\be
J^{\IP^2}_1(\zeta)= {1\over 2} \log \left( -{\rd z \over \rd t} \right) -{1\over 12} \log \left( z^7 \left(1-27 z \right) \right)+ A_1. 
\ee
Using (\ref{gones}), as well as 
\be
2 \pi {\partial^2 J_0^{\IP^2} \over \partial \zeta^2}= {\sqrt{3}} {\rd t_c \over \rd t}, 
\ee
one obtains the following expansion for $\CF_1^{\IP^2}(\lambda)$, 
\be
\ba
\CF_1^{\IP^2}(\lambda)&=-{1\over 12} \log \left( \lambda \hbar \right)+\zeta'(-1)+ \frac{5 \pi ^2 \lambda }{18 \sqrt{3}}-\frac{\pi ^4 \lambda ^2}{486}-\frac{40 \pi
   ^6 \lambda ^3}{2187 \sqrt{3}}+\frac{283 \pi ^8 \lambda ^4}{32805}-\frac{1376 \pi ^{10} \lambda ^5}{177147 \sqrt{3}}\\
   & +\frac{72272 \pi ^{12} \lambda
   ^6}{55801305}+\frac{7936 \pi ^{14} \lambda ^7}{14348907 \sqrt{3}}+\CO\left(\lambda ^8\right), 
   \ea
   \ee
  which again is in precise agreement with what was found in (\ref{exp-p2}). 
  
  There is a further test of the first conjectural equality in (\ref{conj-eq}) which 
  can be made at higher genus, by using the results of \cite{hkr}. Let us denote by $F_g(t)$ the higher genus free energy of local $\IP^2$, 
  as computed in \cite{hkr}, with an extra sign $(-1)^{g-1}$. Then, one has  
  \be
  \label{jgp2}
  J^{\IP^2}_g(\zeta)=(4 \pi^2)^{2g-2} \left\{ F_g(t) - 3^{2-2g} (-1)^{g-1} c_g\right\}, \qquad g \ge 2.  
  \ee
The genus $g$ free energy appearing here, $F_g(t)$, includes the so-called constant map contribution, 
\be
F_g(t)= 3 (-1)^{g-1} c_g+ \CO\left(\re^{-t}\right).
\ee
In our formalism, this contribution comes from the first term in the r.h.s. of (\ref{ags}), while the second term leads to an additional constant in (\ref{jgp2}). 
As mentioned above, the higher genus functions $\CF_g^{\IP^2}(\lambda)$ should be given by the symplectic transformation of the (\ref{jgp2}) to the conifold frame. 
This was done for $F_g(t)$ in \cite{hkr}. The resulting quantities, when expanded around the conifold point, display the singular term in $\lambda^{2-2g}$ appearing in 
(\ref{cone-sing}), plus a constant, and a series starting in $\lambda$, i.e. 
\be
F_g^{\rm con}(t_c)= {B_{2g} \over 2g (2g-2)} t_c^{2-2g} + 3^{2-2g} (-1)^{g-1} c_g+ \CO(t_c). 
\ee
Here, $t_c$ is a local flat coordinate around the conifold, proportional to $\lambda$. The constant term appearing here {\it cancels} exactly 
the second term in the r.h.s. of (\ref{jgp2}). This is precisely what is required by (\ref{cone-sing}). In other words, the function 
$A(\hbar)$, which was conjectured in (\ref{ah-p2}) in order to reproduce the spectral properties of $\rho_{\IP^2}$, is precisely what is needed 
to guarantee the matrix model behavior of $\CF_g^{\IP^2}(\lambda)$ near $\lambda=0$. It is also easy to verify that, after taking into account the appropriate normalizations, 
the expansion of $\CF_2^{(1,1)}(\lambda)$ agrees with the genus two topological string free energy at the conifold point. 

As a final remark, note that the planar free energy $\CF^{\IP^2}_0(\lambda)$ can be also computed at strong 't Hooft coupling, 
and an elementary calculation gives 
\be
\CF^{\IP^2}_0(\lambda)\approx -{4 {\sqrt{\pi}} \over 9} \lambda^{3/2}, \qquad \lambda \gg 1. 
\ee
This is the typical behavior for a theory of M2 branes \cite{kt}, and it agrees with the M-theory limit of the free energy computed in \cite{ghm}. 
The behavior of the matrix model for local $\IP^2$ is therefore very similar to what is found in the ABJM matrix model and its generalizations \cite{dmp,hkpt,mp}.

\subsection{Local $\IF_2$}

The case of local $\IF_2$ is an interesting one, since we have an extra parameter $\xi$ in the operator (\ref{f2}). 
Correspondingly, there is an extra K\"ahler parameter in the geometry, which is usually denoted by $T_B$, and related to $\xi$ by 
\be
\label{xiT}
\xi= 2\cosh\left({T_B \over 2} \right). 
\ee
The perturbative part of the modified grand potential has been determined in \cite{gkmr}
\be
J^{({\rm p})}(\mu,\xi, \hbar)
=\frac{2}{3\pi \hbar }\mu^3 +\left\{ {1\over \hbar} \left( \frac{\pi}{3}-\frac{1}{2\pi} \log^2 \left[ \frac{\xi+\sqrt{\xi^2-4}}{2} \right] \right) -{\hbar \over 12 \pi} \right\} \mu
+A(\hbar, \xi), 
\ee
although the function $A(\hbar, \xi)$ is not known in full generality. The genus $g$ free energy of this geometry, $F_g^{\rm inst}(t, t_B)$, 
which appears in the expressions 
(\ref{jg-par}), is now a function of two variables. The variable $t$, defined in (\ref{tr}), is related to $\zeta$ by 
\be
t= 4 \pi \zeta, 
\ee
since $r=2$ in this geometry. The variable $t_B$ is related to $T_B$ by a rescaling of $2 \pi/\hbar$, and it defines a parameter $m$ by the analogue 
of (\ref{xiT}), 
\be
m= 2\cosh\left({t_B \over 2} \right). 
\ee
The genus zero free energy $F_0(t, t_B)$ can be determined, up to integration constants, by the two equations \cite{bt}
\be\label{dtf}
\ba
 {\partial t \over \partial z}&= -{2\over \pi z {\sqrt{1-4(2+m) z}}} K\left( {16 z \over 4(2+m) z -1} \right), \\
 {\partial^2 F_0 \over \partial z \partial t}&= - {2 \over z {\sqrt{1- 4(m-2) z}}} K \left( {4(m+2) z -1 \over 4(m-2) z -1} \right), 
 \ea
 \ee
where $z$ is the global coordinate corresponding to $t$. 

In order to test the second conjectural equality in (\ref{conj-eq}), we have to consider the theory for $\xi=0$. The corresponding value of $T_B$ is
\be
T_B=\pm \pi \ri. 
\ee
Therefore, the 't Hooft limit of the modified grand potential involves the functions written down in (\ref{mod-js}). 
Since we have to expand $F_g^{\rm inst}(t, t_B)$ around $t_B=0$, we have to consider the topological string theory on local $\IF_2$ for the value 
of the parameter $m=2$. For this value, the theory is identical to the diagonal, local $\IP^1 \times \IP^1$ geometry, and (\ref{dtf}) simplifies to 
 \be
 \ba
  {\partial t \over \partial z}&= -{2\over \pi z}  K\left( 16 z \right), \\
 {\partial^2 F_0 \over \partial z \partial t}&= - {2 \over z } K \left( 1-16z \right). 
 \ea
 \ee
This can be integrated explicitly, and one finds the two periods, 
\be
\ba
\omega_1(z)&= \log (z) +4 z \, _4F_3\left(1,1,\frac{3}{2},\frac{3}{2};2,2,2;16 z\right), \\
\omega_2(z)&= {1\over \pi} G_{3,3}^{3,2}\left(16 z\left|
\begin{array}{c}
 \frac{1}{2},\frac{1}{2},1 \\
 0,0,0
\end{array}
\right.\right)-{2 \pi^2 \over 3},
\ea
\ee
which solve the Picard--Fuchs equation
\be
\left( \theta^3-4z \theta(2\theta+1)^2 \right) \Pi=0. 
\ee
The genus zero free energy depends now only on a single parameter $t$, and we will denote it $F_0(t)= F_0(t, 0)$. It is determined again by (\ref{spec}), 
and one finds the expansion, 
 \be
 F_0(t)= {t^3\over 6}- 4\re^{-t}- {9 \over 2} \re^{-2t}-\cdots
 \ee
Collecting the above results, one finds that the function $\tilde J_0^{\IF_2}(\zeta, \xi=0)$ is given by 
\be
\tilde J_0^{\IF_2}(\zeta, \xi=0)= {1\over 16 \pi^4} \left( F_0(t)- {\pi^2 t \over 3} +16 \pi^4 A_0(\xi=0) \right). 
\ee

\begin{figure}
\center
\includegraphics[height=5cm]{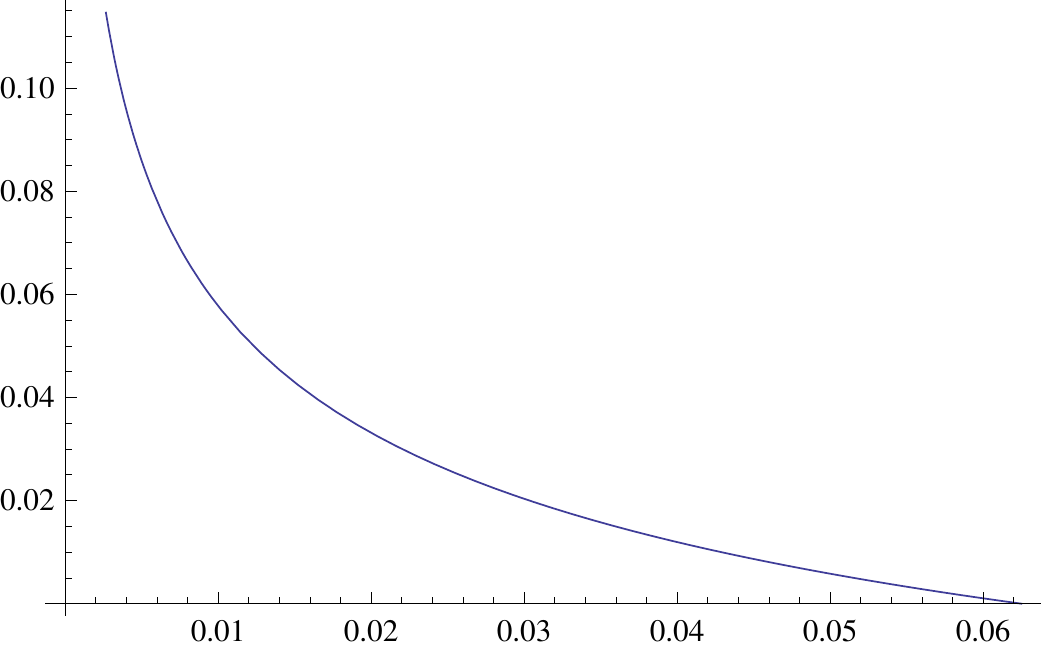} 
\caption{The 't Hooft parameter $\lambda=N/\hbar$ as a function of the modulus $0\le z\le 1/16$. The large radius point $z=0$ corresponds to 
strong 't Hooft coupling, while $z=1/16$, the conifold point, is the weakly coupled theory.}
\label{thooftf2}
\end{figure}

 The equation (\ref{l-jo}) determining the 't Hooft parameter reads in this case, 
 \be
4 \pi^3  \lambda= { \partial F_0(t) \over \partial t} - {\pi^2 \over 3}. 
\ee
As in the case of local $\IP^2$, the r.h.s. of this equation is a vanishing period at the conifold point, located at 
\be
z={1\over 16}. 
\ee
The 't Hooft parameter varies between $0$ and $\infty$ as $z$ varies between $1/16$ and $0$, as shown in \figref{thooftf2}. 
As in the case of local $\IP^2$, we are interested in expanding the quantities at weak 't Hooft coupling, near the conifold point. We introduce the local variable 
\be
y=1-16 z. 
\ee
There is a flat coordinate near the conifold point $y=0$, with the expansion, 
\be
t_c(y)=y+\frac{5 y^2}{8}+\frac{89 y^3}{192}+\frac{381 y^4}{1024}+\frac{25609 y^5}{81920}+\CO\left(y^6\right). 
\ee
It is related to the 't Hooft parameter by 
\be
\lambda={t_c(y) \over 4 \pi^2}. 
\ee
The period $\omega_1(z)$ has the following expression near the conifold point, 
\be
\label{om1C}
\omega_1(z)= -{8 C\over \pi}+ {t_c(y) \over \pi}  \log \left( {y \over 16}\right)+ {1\over \pi} s(y), 
\ee
where 
\be
s(y)=-y-\frac{y^2}{16}+\frac{35 y^3}{288}+\frac{2141 y^4}{12288}+\frac{465061 y^5}{2457600}+\CO\left(y^6\right).
\ee
The genus zero free energy $\CF^{\IF_2}_0(\lambda)$ is determined by the equation
\be
{\partial \CF^{\IF_2}_0(\lambda) \over \partial \lambda}= {\omega_1(z) \over 4 \pi}. 
\ee
This can be integrated to give, up to an integration constant, 
\be
\ba
\CF^{\IF_2}_0(\lambda, \xi=0)&= -{2 C \over \pi^2} \lambda+ \lambda ^2 \left(\frac{\log (\lambda )}{2}-\frac{3}{4}
+\log \left(\frac{\pi }{2}\right)\right) -\frac{\pi ^2 \lambda ^3}{12}+\frac{5 \pi ^4 \lambda
   ^4}{288}-\frac{7 \pi ^6 \lambda ^5}{960}\\
   &+\frac{733 \pi ^8 \lambda ^6}{172800}-\frac{47 \pi ^{10} \lambda ^7}{16128}+\CO\left(\lambda^8\right).
\ea
\ee
This expansion agrees with the result for $\CF^{(2,1)}_0(\lambda)$, as obtained from the 
matrix model in (\ref{exp-f2}). The coefficient in the linear term in $\lambda$ involves the Catalan number, 
due to (\ref{om1C}), and this is precisely what is required to agree with the value of the potential $V^{(0)}_{1,2}(u_\star)$ at its minimum. Full agreement between 
both expressions requires the integration constant for $\CF^{\IF_2}_0(\lambda)$ to vanish. We have verified numerically that this will be the case if 
\be
\label{a0f2}
A_0(\xi=0)= {5 \zeta(3) \over 16 \pi^4}. 
\ee 
It would be interesting to determine the function $A(\hbar, \xi)$ and check whether or not it satisfies (\ref{a0f2})\footnote{While this paper was being typed, 
Y. Hatsuda found an ansatz for the function $A(\hbar, \xi=0)$ \cite{hatsu} which indeed leads to (\ref{a0f2}).}.
 
 Note that $\CF^{\IF_2}_0(\lambda, \xi=0)$ is {\it not} the genus zero free energy of local $\IF_2$ in the conifold frame and with $m=0$. Rather, it corresponds to 
 the value $m=2$. The reason for this ``mismatch" is that, in the conjecture of \cite{ghm}, the K\"ahler parameters corresponding to the mass parameters appear 
 divided by $\hbar$ in the worldsheet instanton expansion. Therefore, when $\hbar \rightarrow \infty$, they are frozen to the values which correspond to $t_{x_i}=0$.

Let us now work out the next-to-leading order in $1/\hbar^2$, which is given by the second line of (\ref{mod-js}). We first note that 
 \be
 {\partial F_0 \over \partial t_B} \biggr|_{t_B=0}= 0, \qquad 
 {\partial^2 F_0 \over \partial t_B^2} \biggr|_{t_B=0}= {1\over 2}{\partial F_0 \over \partial m}\biggr|_{m=2}. 
\ee
This last derivative can be obtained, as a function on moduli space, by taking a further derivative w.r.t. $t$ and using (\ref{dtf}):
\be
{\partial^2 F_0 \over \partial m \partial y}\biggr|_{m=2}={1 \over4 \sqrt{y}} K(y) \, _2F_1\left(\frac{1}{2},\frac{1}{2};2;\frac{y-1}{y}\right)+{1\over \pi  (y-1)} K(1-y) (E(y)-1), 
\ee
where $K(y)$, $E(y)$ are the elliptic integrals of the first and the second kind, respectively, and we have expressed it already in terms of the 
conifold variable. From the general formula in (\ref{mod-js}), one finds
\be
\label{j1f2}
\tilde J^{\IF_2}_1(\zeta)={\pi \zeta \over 8}+A_1+F_1(t, 0)-{1\over 16} {\partial F_0 \over \partial m}\biggr|_{m=2}, 
\ee
where
\be
F_1(t, 0)= -{1\over 2} \log \left( {K (16 z) \over \pi} \right) -{1\over 12} \log\left( 64 z (1-16 z) \right). 
\ee
Using again (\ref{gones}), but this time for the functions $\tilde J^{\IF_2}_{0,1}(\zeta)$, 
we find, 
\be
\label{f2one}
\CF^{\IF_2}_1(\lambda, \xi=0)=-\frac{1}{12} \log (\lambda)+\frac{\pi ^2 \lambda }{6}-\frac{5 \pi ^4 \lambda ^2}{288}+\frac{\pi ^6 \lambda ^3}{576}+\frac{53
   \pi ^8 \lambda ^4}{17280}-\frac{7 \pi ^{10} \lambda ^5}{1152}+\CO\left(\lambda ^6\right), 
   \ee
 up to an additive constant independent of $\lambda$ (but depending on $\hbar$ through $A_1$). This again agrees with the expansion in (\ref{exp-f2}). Note 
 however that the function $\CF^{\IF_2}_1(\lambda, \xi=0)$ is {\it not} the genus one free energy of local $\IF_2$ at the conifold point and for $m=0$ (or 
 even $m=2$), since it involves additional terms. This reflects the fact that the relation between the parameters appearing in the operator and the ``mass" 
 parameters appearing in the geometry depends on $\hbar$. In any case, the result 
 we have obtained is in perfect agreement with the conjecture of \cite{ghm}, and gives a non-trivial test of the way in which 
 mass parameters are incorporated in that framework. 
 
 Although (\ref{f2one}) is not the standard genus one topological string free energy of local $\IF_2$, this does not contradict our claim that one 
 can obtain these free energies from the mirror curve operators and their fermionic traces. If we want to recover the standard dependence on the 
 mass parameters, we have to take a sort of Veneziano limit, and 
 scale the parameters appearing in the operators in an appropriate way, as $\hbar$ grows large. 
 For example, if we consider the fermionic traces $Z(N, \xi, \hbar)$ of the operator $\rho_{\IF_2}$, in the 't Hooft limit (\ref{thooftlimit}), 
 and we choose the following scaling with $\hbar$ for the 
 parameter appearing in the operator,  
 \be
 \label{sc-xi}
 \xi \approx \exp\left( {\hbar t_B \over 4 \pi} \right), 
 \ee
the conjecture of \cite{ghm} says that the resulting $1/N$ expansion would be governed by the 
standard topological string free energy on local $\IF_2$, with a generic K\"ahler parameter $t_B$\footnote{This expectation 
has been checked in detail in \cite{kmz}, after the first version 
of this paper was posted.}.

\sectiono{Conclusions and open problems}

In this paper we have proposed a new type of matrix models whose $1/N$ expansion gives the all-genus free energy of topological 
strings on toric CY threefolds. Our proposal is based on the conjecture of \cite{ghm}, and it is conceptually clear and elegant, as it can be seen in \figref{rels}: given a mirror 
curve, its quantization leads to a trace class operator, as shown in \cite{kas-mar}. The fermionic traces of these operators admit an integral representation, and this leads to 
the matrix models of this paper. This construction provides a non-perturbative completion of the standard topological string, as it was made clear already in \cite{ghm}. However, 
the representation in terms of matrix models spelled out in detail in this paper is particularly appealing, since it involves the standard 't Hooft limit which underlies the string/gauge theory correspondence. 

In practice, in order to study these matrix models, one needs an explicit expression for the kernel of the corresponding 
trace class operator. This was achieved in \cite{kas-mar} in some cases, 
and the resulting matrix models are labelled by two numbers $m$, $n$. They describe topological string theory 
on the anticanonical bundle of the weighted projective space $\IP(1,m,n)$, 
which can be obtained as limits of well-known toric geometries. In particular, the case $m=n=1$ gives an explicit matrix model for local $\IP^2$. 

The fact that the 't Hooft expansion of these matrix models agrees with the topological string free energy is an analytic, non-trivial test of the conjecture of \cite{ghm}, 
which in addition probes the non-perturbative sector of the spectral problem. In this paper we have performed two detailed comparisons, 
involving local $\IP^2$ and local $\IF_2$, and found 
a complete agreement between the matrix model free energies and the topological string free energies. Interestingly, our matrix models  
give the free energies in the conifold frame. It has been suspected for a long time that this is the natural frame for a matrix model 
representation of the topological string, 
due to the conifold behavior of the free energies and the gap condition \cite{hk,hkr}. Our proposal realizes this idea in a very concrete way. Indeed, it {\it implies} 
the gap condition for the CY threefolds that we have studied. The BKMP conjecture \cite{mmopen,bkmp} should be 
also a consequence of our proposal, 
since we give an explicit matrix model realization of the topological string free energies, and very 
likely our matrix models satisfy the topological recursion of \cite{eo} (or a 
variant thereof). 

The matrix models obtained in this paper are very similar to the ones appearing in the localization of Chern--Simons--matter theories. We have now a complete parallelism between 
the theory of Chern--Simons--matter matrix models and the theory of topological strings on toric CY threefolds: in both of them, the perturbative sector is computed by the 
't Hooft expansion of a matrix integral, but there are non-perturbative corrections to the 't Hooft expansion. In the case of the matrix models for topological strings considered here, 
these non-perturbative corrections are explicitly known, since they can be obtained from the modified grand potential of the theory. They correspond to the ``membrane" part of the modified grand potential (\ref{membrane}), which is determined by the NS limit of refined topological string, and they are exponentially small, of the form, 
\be
\exp\left( \hbar \partial_\lambda\CF^S_0(\lambda) \right),
\ee
just like the non-perturbative corrections to the ABJM matrix model obtained in \cite{dmpnp,mp}. 
Note that, since $\hbar \sim 1/g_{\rm st}$, this is indeed non-perturbative in the topological 
string coupling constant.

Clearly, there are many avenues for future research. Our analysis of the matrix models (\ref{zmn-bis}) has been very elementary, and based on a perturbative expansion. 
It would be very interesting to solve these models exactly in the planar limit and beyond, by using for example the techniques of \cite{kostov-exact}. We should also obtain 
explicit representations for the kernels of other operators appearing in the quantization of mirror curves, in order to write down explicit matrix integrals in more 
general cases. Another interesting problem would be to use the topological string free energies and their trans-series extensions, as constructed in for example \cite{cesv1,cesv2} to reconstruct the 
fermionic traces and their matrix model representation via resurgent analysis. In this paper, as in \cite{ghm}, we have 
focused on mirror curves of genus one. It would be important to understand in more detail how to generalize this construction to higher genus mirror curves. 
Finally, it would be interesting to know if the matrix integrals we are writing have some 
gauge theory interpretation, or some underlying M2 brane interpretation, as speculated in \cite{ghm}. 

\section*{Acknowledgements}
We would like to thank Santiago Codesido, Ricardo Couso-Santamar\'\i a, Alba Grassi, Jie Gu, Yasuyuki Hatsuda, Rinat Kashaev, 
Albrecht Klemm, Jonas Reuter and Ricardo Schiappa 
for useful discussions and correspondence. We are particularly thankful to Ricardo Couso-Santamar\'\i a and Ricardo Schiappa for a detailed reading of the draft. 
This work is supported in part by the Fonds National Suisse, 
subsidies 200021-156995 and 200020-141329, and by the NCCR 51NF40-141869 ``The Mathematics of Physics" (SwissMAP).

\appendix
\sectiono{Results for the perturbative expansion}
\label{app-fs}
In this Appendix we list some results for the coefficients $f_{g,k}$ appearing in the weak coupling expansion of the 
free energies in (\ref{planar-f}) and (\ref{cone-sing}), for general $m$, $n$. One finds, for genus zero, 
\footnotesize
\begin{align}
	%f_{0,3}=-\frac{\pi^2}{12(m+n+1)} \frac {\cos\frac{2\pi}{ m+n+1}+\cos \frac{2 \pi m}{m+n+1}+\cos \frac{2 \pi n}{m+n+1}+3} {\sin\frac{\pi}{(m+n+1)} \sin\frac{\pi m}{(m+n+1)}\sin\frac{\pi n}{(m+n+1)}}, \\
	%
	f_{0,3}&=-\frac{\pi^2}{12(m+n+1)} \frac {\sum_{\alpha \in \{1,m,n\}} \cos \frac{2 \pi \alpha}{m+n+1}+3} {\prod_{\alpha \in \{1,m,n\}} \sin \frac{\pi \alpha}{m+n+1}}, \\
	\nonumber
	\\
	f_{0,4}&=\frac{\pi^4}{576(m+n+1)^2} \frac {\sum_{\alpha \in \{1,m,n\}} \Big (110 \cos \frac{2 \pi \alpha}{m+n+1}+\cos \frac{4 \pi \alpha}{m+n+1} \Big )+5\sum_{\alpha,\beta \in \{1,m,n\}}\cos \frac{2 \pi (\alpha-\beta)}{m+n+1}+126} {\prod_{\alpha \in \{1,m,n\}} \sin^2 \frac{\pi \alpha}{m+n+1}}, \\
	\nonumber
	\\
	f_{0,5}&=-\frac{\pi^6}{480(m+n+1)^3} \frac{1}{\prod_{\alpha \in \{1,m,n\}} \sin^3 \frac{\pi \alpha}{m+n+1}} 
	\Big (\sum_{\alpha \in \{1,m,n\}} \Big ( 283\cos \frac{2 \pi \alpha}{m+n+1} +23  \cos \frac{4 \pi \alpha}{m+n+1} \Big) \nonumber \\
	& \qquad \qquad \qquad \qquad  + \sum_{\alpha,\beta \in \{1,m,n\}} \Big ( 30 \cos \frac{2 \pi (\alpha-\beta)}{m+n+1}+\cos \frac{2 \pi (2\alpha-\beta)}{m+n+1} \Big )+183 \Big ).
\end{align}
\normalsize
For genus one, one has
\footnotesize
\begin{align}
		f_{1,1}&=-\frac{1}{2}f_{0,3}+\frac{\pi^2}{4(m+n+1)}\frac{1}{\prod_{\alpha \in \{1,m,n\}} \sin \frac{\pi \alpha}{m+n+1}},
		\\
		f_{1,2}&=-f_{0,4},
		\\
		f_{1,3}&=\frac{\pi^6}{288(m+n+1)^3} \frac{1}{\prod_{\alpha \in \{1,m,n\}} \sin^3 \frac{\pi \alpha}{m+n+1}} 
	\Big (\sum_{\alpha \in \{1,m,n\}} \Big ( 175 \cos \frac{2 \pi \alpha}{m+n+1} +11  \cos \frac{4 \pi \alpha}{m+n+1} \Big) \nonumber \\
	& \qquad \qquad \qquad \qquad  + \sum_{\alpha,\beta \in \{1,m,n\}} \Big ( 18 \cos \frac{2 \pi (\alpha-\beta)}{m+n+1}+\cos \frac{2 \pi (2\alpha-\beta)}{m+n+1} \Big )+99 \Big ).
\end{align}
\normalsize
Finally, for genus two, we obtain:
\footnotesize
\begin{align}
	f_{2,1}&=-\frac{\pi^6}{960(m+n+1)^3} \frac{1}{\prod_{\alpha \in \{1,m,n\}} \sin^3 \frac{\pi \alpha}{m+n+1}} 
	\Big (\sum_{\alpha \in \{1,m,n\}} \Big ( 13\cos \frac{2 \pi \alpha}{m+n+1} -7  \cos \frac{4 \pi \alpha}{m+n+1} \Big) \nonumber \\
	& \qquad \qquad \qquad \qquad  + \sum_{\alpha,\beta \in \{1,m,n\}}\cos \frac{2 \pi (2\alpha-\beta)}{m+n+1} -27 \Big ).\end{align}


\normalsize



\begin{thebibliography}{99}
\bibliographystyle{plain}

\bibitem{dv}
 R.~Dijkgraaf and C.~Vafa, ``Matrix models, topological strings, and supersymmetric gauge theories,''
  Nucl.\ Phys.\ B {\bf 644}, 3 (2002)
  [hep-th/0206255].
  %%CITATION = HEP-TH/0206255;%%
  
 \bibitem{mmcs}
 M.~Mari\~no, ``Chern-Simons theory, matrix integrals, and perturbative three manifold invariants,''
  Commun.\ Math.\ Phys.\  {\bf 253}, 25 (2004)
  [hep-th/0207096].
  %%CITATION = HEP-TH/0207096;%%  
  
 \bibitem{gv} R.~Gopakumar and C.~Vafa, ``On the gauge theory / geometry correspondence,''
  Adv.\ Theor.\ Math.\ Phys.\  {\bf 3}, 1415 (1999)
  [hep-th/9811131].
  %%CITATION = HEP-TH/9811131;%%

\bibitem{akmv}
   M.~Aganagic, A.~Klemm, M.~Mari\~no and C.~Vafa, ``Matrix model as a mirror of Chern-Simons theory,''
  JHEP {\bf 0402}, 010 (2004)
  [hep-th/0211098].
  %%CITATION = HEP-TH/0211098;%%
   
\bibitem{ghm}
 A.~Grassi, Y.~Hatsuda and M.~Mari\~no, ``Topological Strings from Quantum Mechanics,''
  arXiv:1410.3382 [hep-th].
  %%CITATION = ARXIV:1410.3382;%%

\bibitem{adkmv}
 M.~Aganagic, R.~Dijkgraaf, A.~Klemm, M.~Mari\~no and C.~Vafa, ``Topological strings and integrable hierarchies,''
  Commun.\ Math.\ Phys.\  {\bf 261}, 451 (2006)
  [hep-th/0312085].
  %%CITATION = HEP-TH/0312085;%%
  
\bibitem{acdkv}
M.~Aganagic, M.~C.~N.~Cheng, R.~Dijkgraaf, D.~Krefl and C.~Vafa, ``Quantum Geometry of Refined Topological Strings,''
  JHEP {\bf 1211}, 019 (2012)
  [arXiv:1105.0630 [hep-th]].
  %%CITATION = ARXIV:1105.0630;%%

 \bibitem{mirmor}
   A.~Mironov and A.~Morozov, ``Nekrasov Functions and Exact Bohr-Zommerfeld Integrals,''
  JHEP {\bf 1004}, 040 (2010)
  [arXiv:0910.5670 [hep-th]].
  %%CITATION = ARXIV:0910.5670;%%
      
     \bibitem{ns}
   N.~A.~Nekrasov and S.~L.~Shatashvili, ``Quantization of Integrable Systems and Four Dimensional Gauge Theories,''
  arXiv:0908.4052 [hep-th].
  %%CITATION = ARXIV:0908.4052;%%  
 
 \bibitem{kwy}
A.~Kapustin, B.~Willett and I.~Yaakov, ``Exact Results for Wilson Loops in Superconformal Chern-Simons Theories with Matter,''
  JHEP {\bf 1003}, 089 (2010)
  [arXiv:0909.4559 [hep-th]].
  %%CITATION = JHEPA,1003,089;%%  
  
   \bibitem{abjm}
 O.~Aharony, O.~Bergman, D.~L.~Jafferis and J.~Maldacena, ``N=6 superconformal Chern-Simons-matter theories, M2-branes and their gravity duals,''
  JHEP {\bf 0810}, 091 (2008)
  [arXiv:0806.1218 [hep-th]].
  %%CITATION = JHEPA,0810,091;%%
  
\bibitem{mm-lectures}
  M.~Mari\~no, ``Lectures on localization and matrix models in supersymmetric Chern-Simons-matter theories,''
  J.\ Phys.\ A {\bf 44}, 463001 (2011)
  [arXiv:1104.0783 [hep-th]].
  %%CITATION = ARXIV:1104.0783;%%
  


    \bibitem{km}
 J.~Kallen and M.~Mari\~no, ``Instanton effects and quantum spectral curves,''
  arXiv:1308.6485 [hep-th].
  %%CITATION = ARXIV:1308.6485;%%
  
\bibitem{kas-mar}
  R.~Kashaev and M.~Mari\~no, ``Operators from mirror curves and the quantum dilogarithm,''
  arXiv:1501.01014 [hep-th].
  
\bibitem{hw}
 M.~x.~Huang and X.~f.~Wang, ``Topological Strings and Quantum Spectral Problems,''
  JHEP {\bf 1409}, 150 (2014)
  [arXiv:1406.6178 [hep-th]].
  %%CITATION = ARXIV:1406.6178;%%
    
   
 \bibitem{kallen} J.~Kallen, ``The spectral problem of the ABJ Fermi gas,''
  arXiv:1407.0625 [hep-th].
  %%CITATION = ARXIV:1407.0625;%% 
  
 \bibitem{fhw}
  X.~f.~Wang, X.~Wang and M.~x.~Huang, ``A Note on Instanton Effects in ABJM Theory,''
  arXiv:1409.4967 [hep-th].
  %%CITATION = ARXIV:1409.4967;%% 
 
 
\bibitem{zjj}
J.~Zinn-Justin and U.~D.~Jentschura, ``Multi-instantons and exact results I: Conjectures, 
WKB expansions, and instanton interactions,''
  Annals Phys.\  {\bf 313}, 197 (2004)
  [quant-ph/0501136].
  %%CITATION = QUANT-PH/0501136;%%
  
 \bibitem{cgm}
 S.~Codesido, A.~Grassi and M.~Mari\~no, ``Exact results in N=8 Chern-Simons-matter theories and quantum geometry,''
  arXiv:1409.1799 [hep-th].

 
\bibitem{hy}
 N.~Halmagyi and V.~Yasnov, ``The Spectral curve of the lens space matrix model,''
  JHEP {\bf 0911}, 104 (2009)
  [hep-th/0311117].
  %%CITATION = HEP-TH/0311117;%%

\bibitem{hoy}
N.~Halmagyi, T.~Okuda and V.~Yasnov, ``Large N duality, lens spaces and the Chern-Simons matrix model,''
  JHEP {\bf 0404}, 014 (2004)
  [hep-th/0312145].
  %%CITATION = HEP-TH/0312145;%%  
  

  \bibitem{eynard} B.~Eynard, ``All orders asymptotic expansion of large partitions,''
  J.\ Stat.\ Mech.\  {\bf 0807}, P07023 (2008)
  [arXiv:0804.0381 [math-ph]].
  %%CITATION = ARXIV:0804.0381;%%

\bibitem{ks}
A.~Klemm and P.~Sulkowski, ``Seiberg-Witten theory and matrix models,''
  Nucl.\ Phys.\ B {\bf 819}, 400 (2009)
  [arXiv:0810.4944 [hep-th]].
  %%CITATION = ARXIV:0810.4944;%%

  
  \bibitem{sul}
  P.~Sulkowski, ``Matrix models for 2* theories,''
  Phys.\ Rev.\ D {\bf 80}, 086006 (2009)
  [arXiv:0904.3064 [hep-th]].
  %%CITATION = ARXIV:0904.3064;%%
 
 \bibitem{ekm1}
 B.~Eynard, A.~K.~Kashani-Poor and O.~Marchal, ``A Matrix Model for the Topological String I: Deriving the Matrix model,''
  Annales Henri Poincare {\bf 15}, 1867 (2014)
  [arXiv:1003.1737 [hep-th]].
  %%CITATION = ARXIV:1003.1737;%%
  
 \bibitem{ekm2} B.~Eynard, A.~K.~Kashani-Poor and O.~Marchal, 
 ``A Matrix model for the topological string II. The spectral curve and mirror geometry,''
  Annales Henri Poincare {\bf 14}, 119 (2013)
  [arXiv:1007.2194 [hep-th]].
  %%CITATION = ARXIV:1007.2194;%%
   
      \bibitem{simon} 
 B. Simon, {\it Trace ideals and their applications}, second edition, American Mathematical Society, Providence, 2000. 

        \bibitem{zamo}
  A.~B.~Zamolodchikov, ``Painlev\'e III and 2-d polymers,''
  Nucl.\ Phys.\ B {\bf 432}, 427 (1994)
  [hep-th/9409108].
  %%CITATION = HEP-TH/9409108;%%  
   
\bibitem{kostov}  I.~K.~Kostov, ``Solvable statistical models on a random lattice,''
  Nucl.\ Phys.\ Proc.\ Suppl.\  {\bf 45A}, 13 (1996)
  [hep-th/9509124].
  %%CITATION = HEP-TH/9509124;%%
      
      \bibitem{grassi-marino}
 A.~Grassi and M.~Mari\~no, ``M-theoretic matrix models,''
  arXiv:1403.4276 [hep-th].
  %%CITATION = ARXIV:1403.4276;%%  

  \bibitem{hhl}
  N.~Hama, K.~Hosomichi and S.~Lee, ``Notes on SUSY Gauge Theories on Three-Sphere,''
  JHEP {\bf 1103}, 127 (2011)
  [arXiv:1012.3512 [hep-th]].
  %%CITATION = ARXIV:1012.3512;%%
  
  \bibitem{jafferis}
 D.~L.~Jafferis, ``The Exact Superconformal R-Symmetry Extremizes Z,''
  JHEP {\bf 1205}, 159 (2012)
  [arXiv:1012.3210 [hep-th]].
  %%CITATION = ARXIV:1012.3210;%%   

      \bibitem{hkp}
 M.~X.~Huang, A.~Klemm and M.~Poretschkin, ``Refined stable pair invariants for E-, M- and $[p, q]$-strings,''
  JHEP {\bf 1311}, 112 (2013)
  [arXiv:1308.0619 [hep-th]].
  %%CITATION = ARXIV:1308.0619;%%  
  
  \bibitem{hkrs}
  M.~x.~Huang, A.~Klemm, J.~Reuter and M.~Schiereck, ``Quantum geometry of del Pezzo surfaces in the Nekrasov-Shatashvili limit,''
  arXiv:1401.4723 [hep-th].
  %%CITATION = ARXIV:1401.4723;%%
 
  \bibitem{bc}
  A.~Brini and R.~Cavalieri, ``Crepant Resolutions and Open Strings II,''
  arXiv:1407.2571 [math.AG].
  %%CITATION = ARXIV:1407.2571;%% 

\bibitem{faddeev}
L.~D. Faddeev, ``Discrete {H}eisenberg-{W}eyl group and modular group,"
  Lett. Math. Phys. \textbf{34},  249 (1995).

\bibitem{fk}
L.~D. Faddeev and R.~M. Kashaev, ``Quantum dilogarithm," Modern
  Phys. Lett. A \textbf{9}, 427 (1994).

\bibitem{kwy2}
A.~Kapustin, B.~Willett and I.~Yaakov, ``Nonperturbative Tests of Three-Dimensional Dualities,''
  JHEP {\bf 1010}, 013 (2010)
  [arXiv:1003.5694 [hep-th]].
  %%CITATION = ARXIV:1003.5694;%%  

       \bibitem{mp}
 M.~Mari\~no and P.~Putrov, ``ABJM theory as a Fermi gas,''
  J.\ Stat.\ Mech.\  {\bf 1203}, P03001 (2012)
  [arXiv:1110.4066 [hep-th]].
  %%CITATION = ARXIV:1110.4066;%%  


\bibitem{ak}
J.~Ellegaard Andersen and R.~Kashaev, ``A TQFT from Quantum Teichm\"uller Theory,''
  Commun.\ Math.\ Phys.\  {\bf 330}, 887 (2014)
  [arXiv:1109.6295 [math.QA]].
  %%CITATION = ARXIV:1109.6295;%%
  
   \bibitem{kostov-exact}
 I.~K.~Kostov, ``Exact solution of the six vertex model on a random lattice,''
  Nucl.\ Phys.\ B {\bf 575}, 513 (2000)
  [hep-th/9911023].
  %%CITATION = HEP-TH/9911023;%%
  
\bibitem{hmmo}
Y.~Hatsuda, M.~Mari\~no, S.~Moriyama and K.~Okuyama, ``Non-perturbative effects and the refined topological string,''
  JHEP {\bf 1409}, 168 (2014)
  [arXiv:1306.1734 [hep-th]].
  %%CITATION = ARXIV:1306.1734;%%

 
     \bibitem{hmo} 
  Y.~Hatsuda, S.~Moriyama and K.~Okuyama, ``Instanton Effects in ABJM Theory from Fermi Gas Approach,''
  JHEP {\bf 1301}, 158 (2013)
  [arXiv:1211.1251 [hep-th]].
  %%CITATION = ARXIV:1211.1251;%%  
 
  
     \bibitem{kkn}
   V.~A.~Kazakov, I.~K.~Kostov and N.~A.~Nekrasov, ``D particles, matrix integrals and KP hierarchy,''
  Nucl.\ Phys.\ B {\bf 557}, 413 (1999)
  [hep-th/9810035].
  %%CITATION = HEP-TH/9810035;%%
  
    
   \bibitem{abk}
  M.~Aganagic, V.~Bouchard and A.~Klemm, ``Topological Strings and (Almost) Modular Forms,''
  Commun.\ Math.\ Phys.\  {\bf 277}, 771 (2008)
  [hep-th/0607100].
  %%CITATION = HEP-TH/0607100;%% 

\bibitem{gvcone}
D.~Ghoshal and C.~Vafa, ``$c = 1$ string as the topological theory of the conifold,''
  Nucl.\ Phys.\ B {\bf 453}, 121 (1995)
  [hep-th/9506122].
  %%CITATION = HEP-TH/9506122;%%
  

\bibitem{hk}
 M.~x.~Huang and A.~Klemm, ``Holomorphic Anomaly in Gauge Theories and Matrix Models,''
  JHEP {\bf 0709}, 054 (2007)
  [hep-th/0605195].
  %%CITATION = HEP-TH/0605195;%%
  
      \bibitem{hanada}
 M.~Hanada, M.~Honda, Y.~Honma, J.~Nishimura, S.~Shiba and Y.~Yoshida, 
 ``Numerical studies of the ABJM theory for arbitrary N at arbitrary coupling constant,''
  JHEP {\bf 1205}, 121 (2012)
  [arXiv:1202.5300 [hep-th]].
  %%CITATION = ARXIV:1202.5300;%%  

\bibitem{ho}
Y.~Hatsuda and K.~Okuyama, ``Probing non-perturbative effects in M-theory,''
  JHEP {\bf 1410}, 158 (2014)
  [arXiv:1407.3786 [hep-th]].
  %%CITATION = ARXIV:1407.3786;%%
  
\bibitem{hkr}
B.~Haghighat, A.~Klemm and M.~Rauch, ``Integrability of the holomorphic anomaly equations,''
  JHEP {\bf 0810}, 097 (2008)
  [arXiv:0809.1674 [hep-th]].
  %%CITATION = ARXIV:0809.1674;%%
  
  \bibitem{rv}
 F. Rodriguez Villegas, ``Modular Mahler measures, I", in {\it Topics in number theory}, Kluwer Acad. Publ., Dordrecht, 1999, p. 17.
 
  \bibitem{dk}
 C. Doran and M. Kerr, ``Algebraic K-theory of toric hypersurfaces,"  Commun. Number Theory Phys. {\bf 5}, 397 (2011) [arXiv:0809.4669 [math.AG]]. 
 
\bibitem{yang}
 K.~Mohri, Y.~Onjo and S.~K.~Yang, ``Closed submonodromy problems, local mirror symmetry and branes on orbifolds,''
  Rev.\ Math.\ Phys.\  {\bf 13}, 675 (2001)
  [hep-th/0009072].
  %%CITATION = HEP-TH/0009072;%%

  
 \bibitem{kt}
I.~R.~Klebanov and A.~A.~Tseytlin, ``Entropy of near extremal black p-branes,''
  Nucl.\ Phys.\ B {\bf 475}, 164 (1996)
  [hep-th/9604089].
  %%CITATION = HEP-TH/9604089;%% 
 
\bibitem{dmp}
N.~Drukker, M.~Mari\~no and P.~Putrov, ``From weak to strong coupling in ABJM theory,''
  Commun.\ Math.\ Phys.\  {\bf 306}, 511 (2011)
  [arXiv:1007.3837 [hep-th]].
  %%CITATION = ARXIV:1007.3837;%%

    \bibitem{hkpt}
 C.~P.~Herzog, I.~R.~Klebanov, S.~S.~Pufu and T.~Tesileanu,
  ``Multi-Matrix Models and Tri-Sasaki Einstein Spaces,''
  Phys.\ Rev.\ D {\bf 83}, 046001 (2011)
  [arXiv:1011.5487 [hep-th]].
  %%CITATION = ARXIV:1011.5487;%%

\bibitem{gkmr}
J. Gu, A. Klemm, M. Mari\~no and J. Reuter, to appear.   

\bibitem{bt}
A.~Brini and A.~Tanzini, ``Exact results for topological strings on resolved Y**p,q singularities,''
  Commun.\ Math.\ Phys.\  {\bf 289}, 205 (2009)
  [arXiv:0804.2598 [hep-th]].
  %%CITATION = ARXIV:0804.2598;%%

\bibitem{hatsu}
Y.~Hatsuda, ``Spectral zeta function and non-perturbative effects in ABJM Fermi-gas,''
  arXiv:1503.07883 [hep-th].
  %%CITATION = ARXIV:1503.07883;%%
  
   \bibitem{kmz}
  R.~Kashaev, M.~Mari\~no and S.~Zakany, ``Matrix models from operators and topological strings, 2,''
  arXiv:1505.02243 [hep-th].
  %%CITATION = ARXIV:1505.02243;%%
  
\bibitem{mmopen}
 M.~Mari\~no, ``Open string amplitudes and large order behavior in topological string theory,''
  JHEP {\bf 0803}, 060 (2008)
  [hep-th/0612127].
  %%CITATION = HEP-TH/0612127;%%
  
  \bibitem{bkmp}
V.~Bouchard, A.~Klemm, M.~Mari\~no and S.~Pasquetti, ``Remodeling the B-model,''
  Commun.\ Math.\ Phys.\  {\bf 287}, 117 (2009)
  [arXiv:0709.1453 [hep-th]].
  %%CITATION = ARXIV:0709.1453;%%

\bibitem{eo}
 B.~Eynard and N.~Orantin, ``Invariants of algebraic curves and topological expansion,''
  Commun.\ Num.\ Theor.\ Phys.\  {\bf 1}, 347 (2007)
  [math-ph/0702045].
  %%CITATION = MATH-PH/0702045;%%

\bibitem{dmpnp}N.~Drukker, M.~Mari\~no and P.~Putrov, ``Nonperturbative aspects of ABJM theory,''
  JHEP {\bf 1111}, 141 (2011)
  [arXiv:1103.4844 [hep-th]].
  %%CITATION = ARXIV:1103.4844;%%
  
 \bibitem{cesv1}
  R.~Couso-Santamar\'\i a, J.~D.~Edelstein, R.~Schiappa and M.~Vonk,
  ``Resurgent Transseries and the Holomorphic Anomaly,''
  arXiv:1308.1695 [hep-th].
  %%CITATION = ARXIV:1308.1695;%%
  
\bibitem{cesv2}
 R.~Couso-Santamar\'\i a, J.~D.~Edelstein, R.~Schiappa and M.~Vonk, 
 ``Resurgent Transseries and the Holomorphic Anomaly: Nonperturbative Closed Strings in Local $\mathbb{C}\mathbb{P}^2$,''
  arXiv:1407.4821 [hep-th].
  %%CITATION = ARXIV:1407.4821;%%  

 \end{thebibliography}
\end{document}